\documentclass[useAMS,usenatbib,usegraphicx]{mn2e}
\usepackage{times}

\def\ginga{{\em Ginga}}
\def\sax{{\em BeppoSAX}}
\def\xmm{{\em XMM-Newton}}

\def\swift{{\em Swift}}

\def\heao{{\em HEAO}}

\def\wise{{\em WISE}}

\def\suzaku{{\em Suzaku}}
\def\nustar{{\em NuSTAR}}

\def\athena{{\em Athena}}

\def\hst{{\em HST}}

\def\p{$\pm$}
\def\ltsim{\mathrel{\hbox{\rlap{\hbox{\lower4pt\hbox{$\sim$}}}\hbox{$<$}}}}
\def\gtsim{\mathrel{\hbox{\rlap{\hbox{\lower4pt\hbox{$\sim$}}}\hbox{$>$}}}}

\def\Msun{M$_{\odot}$}
\def\Lsun{L$_{\odot}$}
\def\Mbh{$M_{\rm BH}$}
\def\micron{$\mu$m}

\def\fexxv{Fe\,{\sc xxv}}
\def\fexxvi{Fe\,{\sc xxvi}}

\def\oiii{[O\,{\sc iii}]}

\def\l{$\lambda$}

\def\lbol{$L_{\rm Bol}$}
\def\ledd{$L_{\rm Edd}$}

\def\ebv{$E_{\rm B-V}$}
\def\nh{$N_{\rm H}$}

\def\eso565{ESO~565--G019}
\def\ftools{{\sc ftools}}
\def\xspec{{\sc xspec}}
\def\apec{{\sc apec}}
\def\pexrav{{\sc pexrav}}
\def\pexmon{{\sc pexmon}}
\def\mytorus{{\sc mytorus}}
\def\torus{{\sc torus}}
\def\fscatt{$f_{\rm scatt}$}
\def\loiii{$L_{\rm [O\,{\scriptscriptstyle III}]}$}

\def\ltwoten{$L_{\rm 2-10}$}
\def\lmir{$L_{\rm MIR}$}

\def\obj{NGC\,7674}
\def\thetainc{$\theta_{\rm inc}$}
\def\thetator{$\theta_{\rm tor}$}

\def\fek{Fe\,K$\alpha$}
\def\afe{$A_{\rm Fe}$}
\def\rsub{$R_{\rm sub}$}
\def\heasoft{{\sc heasoft}}
\def\caldb{{\sc caldb}}

\title[NuSTAR observations of NGC 7674]{The weak Fe fluorescence line and long-term X-ray evolution of the Compton-thick AGN in NGC\,7674}

\author[P. Gandhi et al.]{P. Gandhi,$^{1,2}$ 
A. Annuar,$^{2}$ G.B. Lansbury,$^{2,3}$ D. Stern,$^{4}$ D.M. Alexander,$^{2}$ F.E. Bauer,$^{5,6,7}$\newauthor S. Bianchi,$^8$ S.E. Boggs,$^9$ P.G. Boorman,$^1$ W.N. Brandt,$^{10, 11, 12}$ M. Brightman,$^{13}$\newauthor F.E. Christensen,$^{14}$ A. Comastri,$^{15}$ W.W. Craig,$^{14,16}$ A. Del Moro,$^{17}$ M. Elvis,$^{18}$\newauthor M. Guainazzi,$^{19, 20}$ C.J. Hailey,$^{21}$ F.A. Harrison,$^{13}$ M. Koss,$^{22}$ I. Lamperti,$^{22}$ G. Malaguti,$^{23}$\newauthor A. Masini,$^{15,24}$ G. Matt,$^8$ S. Puccetti,$^{25,26}$ C. Ricci,$^5$ E. Rivers,$^{13}$ D.J. Walton$^{3,4,13}$,\newauthor W.W. Zhang$^{27}$\\
Author affiliations appear at the end of the paper. 
}

\begin{document}

\date{\today}


\maketitle
\label{firstpage}

\begin{abstract}
We present \nustar\ X-ray observations of the active galactic nucleus (AGN) in \obj. The source shows a flat X-ray spectrum, suggesting that it is obscured by Compton-thick gas columns. Based upon long-term flux dimming, previous work suggested the alternate possibility that the source is a recently switched-off AGN with the observed X-rays being the lagged echo from the torus. Our high-quality data show the source to be reflection-dominated in hard X-rays, but with a relatively weak neutral \fek\ emission line (equivalent width [EW] of $\approx$\,0.4\,keV) and a strong \fexxvi\ ionised line (EW\,$\approx$\,0.2\,keV). We construct an updated long-term X-ray light curve of NGC\,7674 
and find that the observed 2--10 keV flux has remained constant for the past $\approx$\,20 years, following a high flux state probed by \ginga. Light travel time arguments constrain the minimum radius of the reflector to be $\sim$\,3.2\,pc under the switched-off AGN scenario, $\approx$\,30\,times larger than the expected dust sublimation radius, rendering this possibility unlikely. 
A patchy Compton-thick AGN (CTAGN) solution is plausible, requiring a minimum line-of-sight column density (\nh) of 3\,$\times$\,10$^{24}$\,cm$^{-2}$ at present, and yields an intrinsic 2--10\,keV luminosity of (3--5)\,$\times$\,10$^{43}$\,erg\,s$^{-1}$. 
Realistic uncertainties span the range of $\approx$\,(1--13)\,$\times$\,10$^{43}$\,erg\,s$^{-1}$. The source has one of the weakest fluorescence lines amongst {\em bona fide} CTAGN, and is potentially a local analogue of bolometrically luminous systems showing complex neutral and ionised Fe emission. 
It exemplifies the difficulty of identification and proper characterisation of distant CTAGN based on the strength of the neutral \fek\ line. 
\end{abstract}

\section{Introduction}

Obscured active galactic nuclei (AGN) dominate the overall population of AGN in the cosmos, especially the efficiently-accreting sources which power the cosmic hard X-ray background radiation \citep[e.g. ][]{settiwoltjer89, comastri95, g03, gilli07, treister09, ballantyne11, ueda14}. Yet, finding and characterising these objects is made difficult by the strong absorption they can suffer across the electromagnetic spectrum. In particular, the census of the sources hidden behind extreme obscuring column densities of gas -- in particular Compton-thick AGN with column densities \nh\,$\gtsim$\,1.5\,$\times$\,10$^{24}$\,cm$^{-2}$; hereafter, CTAGN -- remains highly incomplete (see \citealt{ricci15} and \citealt{koss16} for recent updates on the hard X-ray selected CTAGN census). Even locally, very few robust ({\em bona fide}) CTAGN are known \citep{dellaceca08, goulding12, g14}. 

One candidate of a nearby, luminous CTAGN is \obj, the brightest member of the Hickson\,96 interacting galaxy group. \obj\ is a known Seyfert 2, showing strong narrow optical emission lines with full widths at half maximum (FWHM) of less than 500\,km\,s$^{-1}$ in spectral observations carried out over three decades ago \citep{feldman82}. The source also shows a prolific outflow, which manifests as a prominent shoulder on the blue wings of all the prominent narrow optical lines. The outflow has been studied in detail using \hst\ spectroscopy by \citet{fischer13}, who find blueshifts of up to $\sim$\,1700\,km\,s$^{-1}$ (and even larger FWHM) along the narrow line region aligned with the jet axis of the source. In the radio, \obj\ shows at least three separate compact components in VLBI observations on scales of $\approx$0.7\,arcsec\ corresponding to $\approx$0.4\,kpc, and there have been suggestions that the collimated ejecta associated with the radio source could also drive the optical-emitting line outflow \citep{unger88}. Even higher resolution VLBI observations by \citet{momjian03} reveal a complex \lq S\rq--shaped structure, which could result from interactions of the jet with the interstellar medium. Using the \oiii\ emission line as a bolometric luminosity indicator, \citet{xu99} classify \obj\ as a radio quiet AGN. 

The source was first reported to be a refection-dominated AGN by \citet{malaguti98} from \sax\ X-ray observations carried out in 1996, with the direct (intrinsic) continuum being fully absorbed by a Compton-thick gas column. \sax\ detected \obj\ over the energy range of $\sim$\,0.1--60\,keV. \citeauthor{malaguti98} reported a complex structure to the neutral \fek\ line, and estimated a high intrinsic AGN luminosity assuming a scattering geometry similar to NGC\,1068. 
\setcounter{footnote}{26}
But the \sax\ observation was not the first X-ray observation of the source. As detailed in a historical X-ray analysis by \citet{bianchi05}, the source was likely detected by the \heao\ mission in the late 1970s at a 2--10\,keV flux level almost 30\,times brighter \citep{grossan92} and subsequently with an intermediate flux by \ginga\ \citep{awaki91}. \citet{bianchi05} discuss several potential caveats to these detections, including \heao\ contamination by nearby sources, and systematic uncertainties related to the background level measured by \ginga, and conclude that the combined weight of evidence favours the historical source detections being real, although the \ginga\ flux measurement is considered to be more reliable of the two, due to a more robust background determination. The \ginga\ spectrum is a lightly absorbed power law (with \nh\,$<$\,2\,$\times$\,10$^{22}$\,cm$^{-2}$) with an upper limit of 80\,eV to the equivalent width (EW) of any neutral \fek\ emission line. The source then declined by an order of magnitude in continuum flux by the time it was observed by \sax\ to be Compton-thick. Such behaviour might argue for the source being a member of the class of \lq changing-look\rq\ AGN \citep[e.g., ][]{matt03}, associated with clumps of obscuring clouds transiting across the line-of-sight (l.o.s) resulting in apparent changes in \nh(l.o.s). The most famous example of this class is NGC\,1365 which shows dramatic \nh\ variability on timescales of days \citep[][]{risaliti05}.

However, an \xmm\ observation carried out 6 years later (in 2004) also showed a reflection-dominated spectrum completely consistent in shape as well as flux with \sax, unlike what may be expected in a changing-look AGN. \citet{bianchi05} interpret the source as potentially having switched-off, with the spectrum observed by \sax\ and \xmm\ being the reflection component which is delayed with respect to the direct, illuminating power law (PL). Another source that has been discussed from these two opposing perspectives recently is NGC\,7582 \citep{rivers15_7582}. \citet{bianchi05} also found a relatively weak \fek\ emission line in \obj, possibly blended with an ionised \fexxvi\ line at 6.97\,keV. 

Approximately 10 years after the last \xmm\ observation, \obj\ was observed by \suzaku\ in 2013 and then by \nustar\ in 2014, in addition to several snapshot observations by the \swift\ satellite between 2011 to 2014. Here, we present a spectral analysis of these unpublished observations, and combine this with historical data to study the long-term source evolution. We discuss and place constraints on the switched-off AGN as well as the CTAGN scenarios. Finally, we touch upon the relevance of the complex Fe lines for the identification and characterisation of distant CTAGN. We assume $H_{\rm 0}$\,=\,67.3\,km\,s$^{-1}$\,Mpc$^{-1}$ and $\Omega_\Lambda$\,=\,0.685 \citep{planckcosmology}, corresponding to a distance of 126\,Mpc corrected to the reference frame defined by the cosmic microwave background. The source systemic redshift is $z$\,=\,0.0289. All X-ray spectral fitting is carried out with the \xspec\ package v12.9.0 \citep{xspec} and uncertainties are quoted at 90\%\ confidence, unless stated otherwise.

\section{Observations}

A log of the \nustar, \suzaku\ and \swift\ X-ray observations analysed herein is presented in Table\,\ref{tab:obslog}, and the individual data sets are described in this section. 

\begin{table}
\begin{center}
\caption{Observation Log}\label{tab:obslog}
\begin{tabular}{lccr}
\hline
\hline
Mission & Instrument(s)  &    Observation date  & Exposure  \\
        &                &              &   ks      \\
\hline
\nustar & FPMA/B         &   2014-09-30   & 52.0/51.9 \\
\suzaku & XIS0/1/3       &   2013-12-08   & 52.2/52.2/52.2 \\
\swift  & XRT            &   2011-01-28...2014-10-08$^\dag$   & 48.8$^\dag$ \\
\hline
\end{tabular}
~\par
$^\dag$\,For \swift, 17 observations are combined here. See Appendix for details.
\end{center}
\end{table}

\subsection{\nustar}

NGC\,7674 was observed by \nustar\ \citep{nustar} for about 52\,ks of exposure on 2014 Sep 30 (ObsID 708023010). \nustar\ is the first orbiting telescope capable of focusing X-rays above $\sim$\,10\,keV, operating over the energy range of 3--79\,keV. The data were processed using standard steps and the \nustar\ Data Analysis Software ({\sc nustardas}) v.1.3.0 which is provided as part of \heasoft\footnote{{\tt https://heasarc.gsfc.nasa.gov}} and associated \ftools\ \citep{ftools}. \nustar\ \caldb\ calibration files were used to generate cleaned event files after filtering for South Atlantic Anomaly passages and the standard depth cut, in order to reduce instrumental background. 

Source spectra were extracted using a circular aperture 45\arcsec\ in radius centred on the source position in both focal plane modules (FPMs). Background spectra were extracted from source free regions on the detector. The {\tt nuproducts} task was used to extract these calibrated spectra and to generate corresponding response files. All spectra were grouped to a minimum signal-to-noise of at least 4 per grouped energy bin after background subtraction for fitting purposes.

The source is well detected in both FPMs with 3--78\,keV count rates per second of 2.92\,\p\,0.08 (FPMA) and 2.53\,\p\,0.08 (FPMB).

\subsection{\suzaku}

About 52\,ks of exposure were obtained on 2013 Dec 08 with \suzaku. The X-ray Imaging Spectrometer (XIS; \citealt{xis}) is sensitive over $\approx$\,0.5--10\,keV. Standard \ftools\ software for \suzaku\ was used for data reduction and cleaned event file generation with recommended filtering. Source counts were extracted from within a 3.4\,arcmin radius aperture for integrating XIS source counts, and background counts from a larger source-free polygon. The generated spectra and responses of the two front-illuminated (FI) detectors were combined together, and this was analysed simultaneously with the back-illuminated (BI) detector data. For the fitting, we ignore the energy ranges of 1.7--1.9\,keV and 2.1--2.3\,keV because of calibration uncertainties.

The source was also observed by the Hard X-ray Detector (HXD; \citealt{hxd, hxdinorbit}). The HXD/PIN data (the PIN array is sensitive between $\approx$\,15--60 keV) were reduced using standard tasks. The \ftools\ routine {\tt hxdpinxbpi} is a pipeline task that first extracts spectral counts and corrects these for deadtime. Using the \lq tuned\rq\ background model for the target observation provided by the \suzaku\ team as a starting point \citep{fukazawa09}, {\tt hxdpinxbpi} also outputs a background spectrum including the cosmic X-ray background (CXB) component. However, after background subtraction, the residual source count rate was found to be 0.013\,cts\,s$^{-1}$ (15--60 keV), which is $\approx$\,5.0\,\% of the gross count rate and similar to the level of background reproducibility for observations after 2012.\footnote{{\tt www.astro.isas.jaxa.jp/suzaku/analysis/hxd/pinnxb/\newline\,tuned/140530bgdd.pdf}} We conservatively consider the source as a non-detection but note that fitting a PL to the detected net counts between 15 and 60 keV returns an observed flux $F_{15-60}$\,$\approx$\,5\,$\times$\,10$^{-12}$\,erg\,s$^{-1}$\,cm$^{-2}$, which is similar to the observed \nustar\ flux in the same band with the best fit models that we will discuss in the Results section. 

\obj\ is also too faint to be detectable in the HXD/GSO array sensitive to much higher energies, and those data are not considered here.

\subsection{\swift}

The source has been observed by \swift\ \citep{swift} on 17 occasions 2011 onwards, with exposure times ranging over $\approx$\,470--5100\,s using the X-Ray Telescope (XRT; \citealt{xrt}) sensitive to photons between $\sim$0.3--10\,keV. The individual observations are listed in the Appendix. We extracted source and background spectra, together with response files, using the standard XRT Data Analysis Software tools with \heasoft. The version of {\sc xrtpipeline} used was 0.13.0. Source counts were extracted within a 20\,\arcsec\ radius aperture, and background was extracted from an off-source sky region. Upper limits (for detection significance less than 10$^{-3}$) were estimated using the {\tt sosta} command in the {\sc ximage} package. 

We first analysed the spectra individually, but the source lies at the limit of detectability in these observations, yielding only weak detections for observations longer than 2\,ks, and non-detections in other cases. Nevertheless, the individual observations (detections and limits) allow a first check for any strong variations with time. These fits are also described in the Appendix, and no significant variations are found. 

We then extracted a coadded XRT spectrum by combining the event files of the individual observations. This yields a dataset with a total exposure time of 48.8\,ks. The source is well detected in this combined exposure, with a net count rate of 
7.9\,(\p\,0.4)\,$\times$\,10$^{-3}$\,ct\,s$^{-1}$ over the energy range of 0.5--10\,keV. 

The source is classified as a non-detection by the Burst Alert Telescope (BAT; \citealt{bat}) sensitive over 14--195\,keV using standard analysis adopted for the 70-month all-sky survey \citep{baumgartner13}, and we do not consider the BAT data further in this work. We do note, however, that a custom analysis by \citet{koss13} finds a 4.2\,$\sigma$ detection at the position of NGC\,7674 with a flux $F_{14-195}$\,=\,9.9$_{-2.4}^{+5.1}$\,$\times$\,10$^{-12}$\,erg\,s$^{-1}$\,cm$^{-2}$, consistent with that inferred from our \nustar\ analysis described later. 

\subsection{Optical Spectroscopy}

In preparation for, and in support of, the \nustar\ observations, we also obtained optical spectroscopy of NGC\,7674 using the Low Resolution Imaging Spectrometer (LRIS) on the Keck Telescope \citep{lris}. The observations were carried out on 2014\,June\,25 through a 1\farcs 0 wide slit, using both the blue (600 lines\,mm$^{-1}$) grism and the red (400 lines\,mm$^{-1}$) grating, with the D560 dichroic. The night was photometric, with seeing close to 0\farcs 7. 

The blue spectral region is dominated by strong emission lines with blueshifted components, as reported in many previous works (e.g. \citealt{feldman82}). The red spectral region contains the isolated Ca\,{\sc ii} absorption triplet, which can be used to estimate the central black hole mass. This estimate is presented in the Appendix. The instrumental resolution in the red spectral region was measured to be 7.4\,\AA\ (FWHM) using arc lamp spectra, corresponding to a velocity resolution of $\sigma_{\rm instr.}$\,=\,107\,km\,s$^{-1}$ close to the observed wavelength of the Ca triplet feature.

\subsection{X-ray Spectral Analysis Methodology}

In this section, we start by checking for consistency of the data between the various missions, and then describe the details of the spectral analysis models. 

\subsubsection{Basic characterisation}

\begin{figure*}
  \begin{center}
   \includegraphics[angle=270,width=12.5cm]{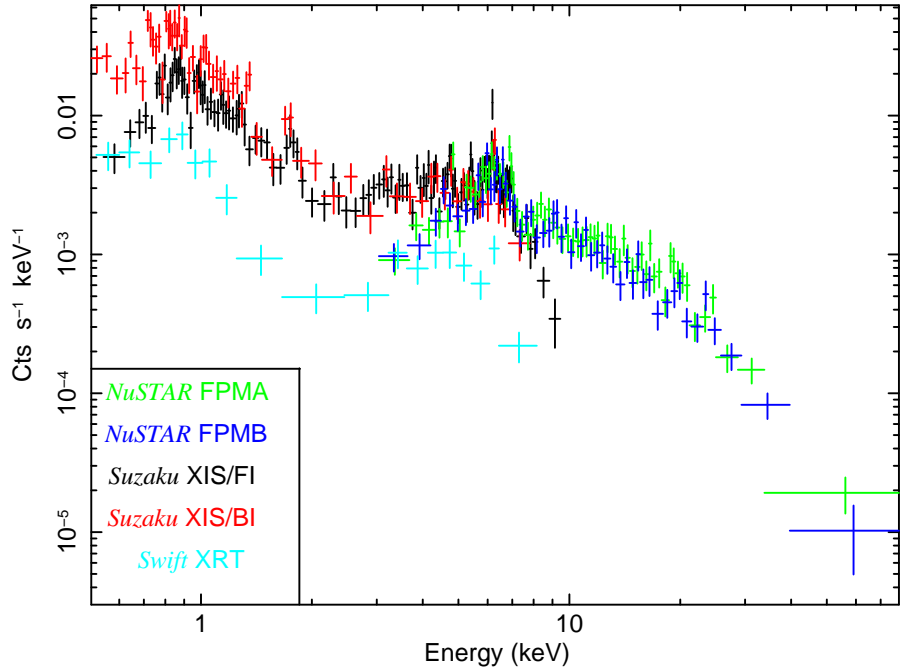}
\caption{The \nustar, \suzaku\ and \swift\ data plotted in observed count rate units. 
 \label{fig:basic}}
  \end{center}
\end{figure*}

Fig.\,\ref{fig:basic} shows the \nustar, \suzaku, and \swift\ data sets overplotted in count rate units, stretching over two decades in observed energy from 0.5--78\,keV. The spectral shape approximately matches between the missions and instruments over common energy ranges. In particular, there appears to be a broad hump dominating the \nustar\ band above 10\,keV and a sharp emission feature around 6\,keV. These are strongly reminiscent of reflection resulting from Compton scattering and neutral Fe fluorescence, a common characteristic of obscured AGN X-ray spectra. The hump extends down to $\sim$\,3\,keV in all data sets, below which a different component appears to dominate in both the \suzaku/XIS and \swift/XRT data with the spectrum rising and peaking around 1\,keV (this is related to the peak in the effective area, in the spectral units of Fig.\,1.).

Fitting a PL to the continuum over the common energy range of 3--10\,keV (after ignoring the range of 5.5--7\,keV around the neutral \fek\ line) simultaneously to all missions returns a photon index $\Gamma$\,=\,0.73\,\p\,0.11 with an acceptable fit statistic of $\chi^2$/dof\,=\,99.9/92. This is much harder than the typical intrinsic photon indices ($\langle\Gamma\rangle\sim 1.9$) seen in AGN X-ray spectra \citep[cf. ][]{nandra97, mateos05_wide, piconcelli05} and is suggestive of heavy obscuration. Fitting the same model to the data from each mission separately, the observed 2--10 keV fluxes span the range of $F_{2-10}$\,$\approx$\,(7--14)\,$\times$\,10$^{-13}$\,erg\,s$^{-1}$\,cm$^{-2}$ between the missions (with the harder \swift\ photon index giving the highest flux), and are fully consistent with each other at 90\,\% confidence. 

We note a mild discrepancy in the \swift\ XRT data with respect to the other missions. When examining the individual photon indices in the above fit to each mission separately, we find $\Gamma_{\rm XRT}$\,=\,--0.68\,\p\,1.26. This is harder than the median $\Gamma$ value from the other missions at 90\,\% confidence. But at 95\% confidence, we find that all missions do agree. The cause of this slight mismatch is not clear but is unlikely to be related to differing aperture sizes used for extracting spectra for the different missions, because we expect the unresolved AGN alone to be the dominant contributor at these energies. Instead, at soft energies, one may expect spatially extended emission and larger differences, which we will discuss later. 
In any case, since this discrepancy is relatively mild, and since all other observations are fully consistent with each other, we consider a joint analysis of the \nustar, \suzaku\ XIS, and coadded \swift\ XRT data to be justified, and this is the approach we follow in the rest of this paper. %





\subsubsection{Reflection models}

For the detailed spectral fits, we will be fitting X-ray reflection models over much of the hard X-ray energy range. There are several canonical models available for fitting heavily obscured AGN spectra in \xspec. Traditionally, \pexrav\ \citep{pexrav} and its successor \pexmon\ including fluorescence \citep{pexmon} have been used to characterise reflection features. These assume a slab obscurer/reflector with an infinite optical depth, as may be expected in a standard geometrically-thin accretion disc. The incident source in X-rays is associated with power law radiation ({PL}$_{\rm AGN}$) from a hot electron accretion disc corona. More recently, there are models which simulate X-ray processing in finite optical depth toroidal media which are more physically appropriate for obscured AGN. \citet{mytorus} provide tabulated results of Monte Carlo simulations of an AGN illuminating a doughnut shaped torus with a fixed opening angle and covering factor of 0.5 (the \mytorus\ model). \citet{brightmannandra11} assume, instead, a torus defined as a conical section of a sphere with variable opening angle and hence variable covering factor (the \torus\ model). Both models assume Solar abundances and treat absorption, reflection and \fek\ fluorescence self-consistently. \mytorus\ additionally allows the freedom to vary the parameters of the scatterer, the l.o.s obscurer, and the fluorescer, decoupled from one another. Both torus models assume that there is no $e$-folding cut-off energy to {PL}$_{\rm AGN}$. For consistency, we make the same assumption in \pexmon. All models assume a uniform gas density spatial distribution. 

Since the geometry of the obscuring/reflecting medium is unknown, we will use all three geometries above and investigate the range in intrinsic properties that can satisfy the observations.


\subsubsection{Additional model components}
\label{sec:additional}

Fixed Galactic absorption ({\sc phabs}$_{\rm Gal}$) with \nh\,=\,4.27\,$\times$\,10$^{20}$\,cm$^{-2}$ \citep{dickeylongman90} was included in all models. We also included constants to account for cross-calibration uncertainties between the various detectors and missions. We found that one or two \apec\ \citep{apec} components and a soft power law ({PL}$_{\rm soft}$) were required to represent the energy range below $\sim$\,2\,keV in some models. We emphasise that these components are meant to serve as parametrisations only. Previous works have shown that the origin of the soft X-ray photons in obscured AGN is a complex mixture of AGN photoionisation, starburst emission, and power law contributions from X-ray binaries, among various possible origins \citep[e.g. ][]{sako00, kinkhabwala02, cappi06, guainazzi07}. Separating these various possible components in \obj\ will require high spatial and spectral resolution observations with {\em Chandra} and \athena, respectively. Our main focus is the origin of the higher energy X-rays, so we use the \apec\ and PL$_{\rm soft}$ components simply to ensure that the spectral fits over the soft regime are statistically acceptable. 
In addition, we may expect spatially extended soft emission to contribute in differing amounts to the \suzaku\ and \swift\ spectra because of the differing spectral extraction apertures tuned to the sizes of the respective telescope point spread functions. We account for this simply by allowing the {\sc apec} components to vary independently between these two missions. The only cross-check required in this regard is that the soft X-ray flux measured by \swift\ XRT (with the smaller aperture) should not exceed that observed in \suzaku\ XIS. This cross-check was applied post-fitting and found to hold for the best fits presented in the following sections. 

With regard to other components, some models preferred several layers of absorption in addition to Galactic, as follows: 

\begin{enumerate}
\item {{\textsl{Nuclear obscuration:}} All models with a transmission component required a thick nuclear absorber, which we associate with the classical compact torus having a l.o.s. column density \lq \nh(nuc)\rq\ well above 10$^{24}$\,cm$^{-2}$.}
\item {{\textsl{Host galaxy absorption:}} Most models preferred the inclusion of a weak absorber screening the soft thermal and power law components (denoted by \lq \nh(host)\rq\ with values of a few times 10$^{21}$\,cm$^{-2}$). This corresponds to weak host galaxy reddening and is consistent with optical reddening of the Narrow Line Region, as we discuss later.} 
\item {{\textsl{Scattering screen:} One alternative scenario that we will investigate requires an additional absorbed power law continuum component, which can potentially be attributed to scattering of the intrinsic PL$_{\rm AGN}$ into the l.o.s. The scattered fraction is $f_{\rm scatt}$, i.e. PL$_{\rm scatt}$\,=\,$f_{\rm scatt}$\,$\times$\,PL$_{\rm AGN}$. Absorption is required for this component, but as we will discuss later, the data cannot distinguish between the two possibilities of (a) a screen that absorbs only the scattered power law, and (b) a screen that affects all the compact nuclear components (direct, reflected, as well as scattered), with both possibilities allowed for in our analysis. We follow the more generic case (b) above, and term this component \lq\nh(scatt).\rq\ It shows intermediate column density values with \nh(scatt)\,$\sim$\,10$^{23}$\,cm$^{-2}$.}} 

\end{enumerate}

\noindent
Finally, we found that inclusion of a Hydrogen-like \fexxvi\ line at 6.97\,keV, probably related to the ionised scattered continuum, provided a significant improvement in all models. 

\vspace*{1cm}
\noindent
Our model values at any energy $E$ can most generically be described as follows. 


\begin{eqnarray}
F(E)\,~~~~~~~~~~~~=\,~~~~~~~~~~~~~C\,e^{-\tau({\rm Gal})}\,e^{-\tau({\rm host})}\,[\,\textrm{\textsc{apec}(s)} + \textrm{PL}_{\rm soft}\nonumber\\
+\,e^{-\tau({\rm {{scatt}}})}\,[\,{\rm Torus}(\Gamma,\,N_{\rm H}({\rm nuc}))\,+\,\textrm{PL}_{\rm {{scatt}}}(\Gamma)\,+\,{\rm Line}\,]\,],\nonumber
\end{eqnarray}

\noindent
where \lq Torus\rq\ represents one of our three primary models including either (i) {\sc pexmon} (Model P), or (ii) \citeauthor{brightmannandra11} \torus\ (Model T), or (iii) \citeauthor{mytorus} \mytorus\ (Model M), in order to model the primary nuclear obscurer and reflector with column density \nh(nuc). \lq $C$\rq\ represents cross-calibration constants, $\tau$ is the optical depth \nh\,$\times$\,$\sigma(E_z)$ at the rest-frame energy $E_z$ of the absorber, and \lq Line\rq\ refers to the \fexxvi\ emission line. We emphasise that not all models require the complexity implied by the above generic description of components. 

\section{Results}





\subsection{Model P: \pexmon}
\label{sec:modelP}

\begin{figure*}
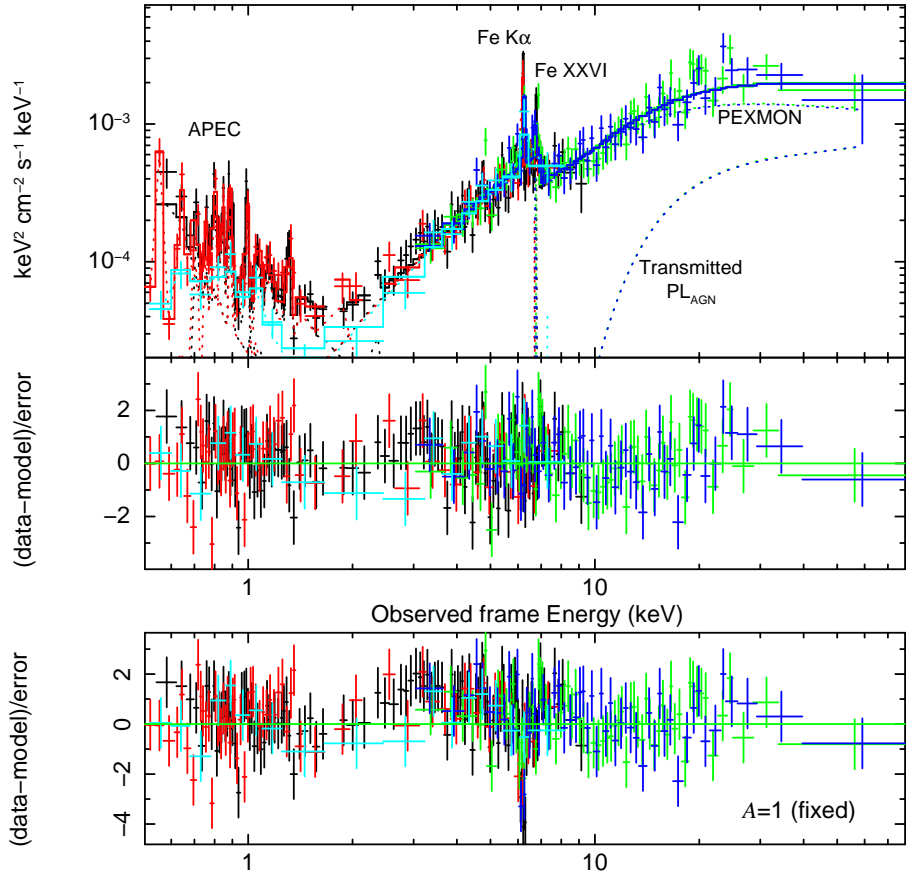

  \begin{center}
    \includegraphics[angle=270,width=12.5cm]{f2a.ps}    
    \includegraphics[angle=270,width=12.5cm]{f2b.ps}
\caption{\pexmon\ model fit to the \suzaku\ XIS FI (black), XIS BI (red), \nustar\ FPMA (green), FPMB (blue) and \swift\ XRT (cyan) data. The central panel shows the fit residuals. 
The Fe abundance for this fit is close to 0.5 
(Table\,\ref{tab:x}). The bottom panel shows the residuals to a fit with fixed abundance $A$\,=\,1. Notice the strong over-prediction at the neutral \fek\ energy. 
 \label{fig:pexmon}}
  \end{center}
\end{figure*}

We began with fitting Model P (\pexmon). This slab reflection model can fit the Compton hump successfully but strongly overpredicts the \fek\ line strength when using fixed abundances at the Solar value. Letting the elemental abundance ($A$\,=\,\afe) vary freely yields an acceptable solution ($\chi^2$/dof\,=\,324.7/305) with $\Gamma$\,=2.1$_{-0.1}^{+0.2}$, $A$\,=\,0.5\,\p\,0.1 and a slab inclination of at least 77\,deg. The inclination affects the \lq peakiness\rq\ of the Compton hump, with lower inclinations being too \lq peaky\rq\ as compared to the data.\footnote{For an illustration of this effect, see Fig.\,5 of \citet{pexrav}.} This model P solution is plotted in Fig.\,\ref{fig:pexmon} and fit parameters are listed in Table\,\ref{tab:x}. The bottom panel in the same figure demonstrates the strong \fek\ residuals with $A$ fixed to Solar and $\Gamma$\,=\,2.1 (fixed to the same value as the canonical model P); letting $\Gamma$ vary freely did not provide a better fit than that stated in Table\,\ref{tab:x}.

A transmitted component of the direct AGN power law ({PL}$_{\rm AGN}$) is included in this model. Compton scattering and photoelectric absorption by the torus are simulated with standard multiplicative models {\sc cabs} and  {\sc zphabs}, respectively, with the gas column density \nh(nuc) tied between the two models. The best-fit \nh(nuc)\,=\,3.4$_{-0.6}^{+0.8}$\,$\times$\,10$^{24}$\,cm$^{-2}$. The reflection component dominates the absorbed transmitted component over the entire spectral range probed, i.e. the source is fully reflection-dominated. The excess of the reflection component is a factor of $\sim$\,3 around 30\,keV. We note that excluding the transmission component results in a moderate increase in the fit statistic to $\chi^2$/dof\,=\,334.5/306, which is only marginally significant at the 3\,$\sigma$ level. In other words, a transmission component is not strongly required. 

We find that the cross-calibration constants between the various missions are consistent with 1 within the uncertainties. The strength of the \fexxvi\ line is EW(\fexxvi)\,$\approx$\,200\,eV. At the soft end, the \suzaku\ XIS data require two \apec\ components with temperatures $kT$\,$\approx$\,0.1 and 0.6\,keV, whereas the \swift\ XRT data probing smaller apertures require only a single lower temperature component. The \apec\ abundances are $\sim$\,0.01--0.2. Multiple thermal models with widely differing temperatures and abundances are quite common in luminous infrared galaxies (e.g. \citealt{ranalli08}) and have also been seen in other CTAGN \citep[e.g. ][]{konami12}. There is, however, strong degeneracy between the abundances and temperatures in data with low spatial and spectral resolution, and tieing all the \suzaku\ and \swift\ model abundances to each other also yields an acceptable solution ($\chi^2$/dof\,=\,332/307) with a global abundance $A$\,=\,0.27$_{-0.14}^{+0.56}$, a \swift\ \apec$_3$ component temperature of $kT_3$\,=\,0.19$_{-0.08}^{+0.03}$\,keV, and no significant changes to the corresponding temperatures for the \apec$_1$ and \apec$_2$ \suzaku\ components. However, we emphasise that we are not attempting to constrain the origin of the soft X-ray emission in detail here. 

The best fit model needs only one additional layer of absorbing column density \nh(host)\,=\,4.0$_{-1.6}^{+2.1}$\,$\times$\,10$^{21}$\,cm$^{-2}$. This is a rather thin screen and corresponds to an optical extinction \ebv\,=\,0.7$_{-0.3}^{+0.4}$\,mag for a standard Galactic gas-to-dust ratio \citep{bohlin78}. This is consistent with the reddening expected from the Balmer decrement of 4.80 \citep{bassani99}. Removing this layer significantly worsens the fit to $\chi^2$/dof\,=\,342.1/306. But most of the change is concentrated at the softest energies (resulting in an increased \apec\ temperature) which we are not modelling in detail; the primary AGN reflection component is unaffected. 




\subsection{Model T: \torus}
\label{sec:modelT}

\pexmon\ assumes a slab geometry with infinite density, which is unlikely to be representative of the toroidal obscurer envisaged by AGN unification schemes. So we next turned to the \citeauthor{brightmannandra11} \torus\ model. This also allows us to explore alternatives to our low \afe\ solution, because abundances are fixed at Solar values in this model. 
In this case, an acceptable solution may be found if we effectively dilute the continuum emission from the torus at the \fek\ line energy with a stronger scattered component (PL$_{\rm scatt}$).\footnote{See the presentation by T. Yaboob at {\tt http://cxc.cfa.harvard.edu/ChandraDecade/} for more discussions of such a warm scatterer scenario.} Such a solution is plotted in Fig.\,\ref{fig:modelT}, where \nh(nuc) is above 10$^{25}$\,cm$^{-2}$, well within the Compton-thick regime. The scattered fraction is \fscatt\,$\approx$\,4\,\%, which raises the continuum at the \fek\ line energy sufficiently to produce a fully acceptable fit ($\chi^2$/dof\,=\,320.2/300). For this fit, we froze the inclination angle \thetainc\ of the torus to be close to edge-on, as is recommended for exploration of the full range of possible covering factors \citep[][]{brightman15}. There are uncertainties in the {\sc torus} model at edge-on inclinations, as discussed by \citet{liu15}. We checked that varying $\theta_{\rm inc}$ did not affect our final inferences. 
The best fit opening angle is \thetator\,=\,62$_{-11}^{+15}$\,deg. Smaller opening angles produce a worse fit because they result in Compton humps peakier than required by the data.\footnote{see Fig.\,1 of \citet{brightman15} for an illustration of the effect of changing \thetator.} Large values of \thetator\ are allowed up to \thetator\,=\,77\,deg, beyond which the torus effectively becomes too geometrically thin to produce a sufficiently strong Compton hump. 

In this model, however, an additional intermediate column obscurer \nh(scatt)\,$\sim$\,10$^{23}$\,cm$^{-2}$ is required, so as not to overproduce the soft X-ray flux. We note that in our model configuration, \nh(scatt)\ affects PL$_{\rm scatt}$ as well as the compact nuclear {\sc torus} component. But modifying the model such that \nh(scatt) affects PL(scatt) alone also yields a fully acceptable solution with $\chi^2$/dof\,=\,323.3/300. This consistency results from the fact that PL$_{\rm scatt}$ completely dominates over the {\sc torus} reflection continuum at energies of around a few keV -- energies below which gas with an intermediate column density of $\sim$\,10$^{23}$\,cm$^{-2}$ is an effective absorber (cf. Fig.\,\ref{fig:modelT}).

This above \nh(scatt) column is $\approx$\,20\,times higher than the host galaxy absorbing layer (\nh(host)) discussed in the previous sub-section, and would imply correspondingly higher optical reddening. So how viable is this? \obj\ is a known luminous infrared galaxy with ongoing star formation and physical interactions with its neighbouring galaxies -- all of which could result in obscuring matter being strewn around the host galaxy on multiple scales. Other examples of nearby galaxies with complex multiple layers of absorption identified by \nustar\ include NGC\,7582 \citep{rivers15_7582} and IC\,751 \citep{ricci16_ic751}. Therefore, the presence of an additional absorber in the innermost parts of the host galaxy which obscures the emergent flux from the nuclear regions is not implausible. 

Detailed spatially resolved studies of the nuclear region so far neither require, nor rule out, such a screen. For example, the study of \citet{fischer13} does not show an obvious \oiii\ flux decrement at the nucleus, but the spatial resolution of their spectral sampling is $\approx$\,30\,pc, much larger than the typical sizes of compact tori, and may potentially allow for additional absorbers below the resolution limit. Alternatively, the outflowing gas associated with the powerful nuclear outflow known in NGC\,7674 may also serve as a screening medium for X-rays. 

In summary, the introduction of \nh(scatt) is clearly somewhat ad hoc and is motivated by the adopted scattering scenario in a fixed Solar metallicity torus model. Nevertheless, we cannot rule it out based upon the present evidence, and we also note that the presence of a PL$_{\rm scatt}$ component, itself, is not surprising, with the measured \fscatt\ value of a few per cent being fully consistent with that seen in other nearby AGN \citep[e.g. ][]{cappi06}. We have introduced this only as one possible scenario, and as we will discuss in the following section, more complex models could alleviate the need for this screen altogether.

The soft energies can be fitted in a very similar fashion to Model P, with one difference being the presence of a significant, but faint, PL$_{\rm soft}$ component. PL$_{\rm soft}$ dominates only in a very narrow energy range around 2\,keV, as a result of which we found that its photon index needed to be fixed, otherwise the fit attempted a very hard slope with contribution to the highest energy \nustar\ range. This component is not required in model P because the reflected component is much more prominent in that case, leaving no deficit around 2\,keV (cf. Fig.\,\ref{fig:pexmon}). Since the purpose of this component is to account for the soft energies, we fixed $\Gamma_{\rm soft}$\,=\,2.0, similar to the spectral slope of the emission from X-ray binaries \citep[e.g. ][]{ranalli03}.  We also checked that allowing small variations of $\Delta$\,$\Gamma_{\rm soft}$\,=\,\p\,0.3 do not affect the modelling of the main AGN component. 

\subsection{Model M: \mytorus}
\label{sec:modelM}

We first fitted a standard \lq coupled\rq\ \mytorus\ model here with all parameters between the l.o.s. and toroidal scattering components tied to each other. 
This solution is very similar in essence to model T, and the fit statistic of $\chi^2$/dof\,=\,325/300 is compatible with both previous models. Quantitatively, models T and M differ in that the derived equatorial column density through the torus (\nh(eq)) is pegged at the allowed upper model threshold of 10$^{25}$\,cm$^{-2}$ in model M. Similarly, \thetainc\ is just above (and very close to) the allowed lower model threshold of \thetainc\,=\,60$^\circ$. Such parameter pegging has been seen in other objects also, and reflects the constraint of the assumed doughnut geometry in this model as a result of which the value of \nh(nuc) is directly coupled to \thetainc\ \citep[e.g., ][]{g14, balokovic14, lansbury15}. Smaller inclination angles are not obscured by the torus in this model, whereas higher \thetainc\ values produce a Compton hump which is too peaky relative to the data. Additional model components are also very similar to model T. The fitted parameters of this model are listed in Table\,\ref{tab:x} and the solution is shown in Appendix figure\,A1. 

\subsubsection{Decoupled \mytorus\ model}
\label{sec:decoupled}

We next tried several versions of the more complex \lq decoupled\rq\ mode in \mytorus. In this mode, the direct l.o.s. absorption, the toroidal Compton scattering emission, and the fluorescence emission components are not necessarily coupled to each other. By varying the \nh, the inclination angles or the relative normalisations associated with these components, one can effectively simulate a variety of scenarios including a clumpy obscurer or varying elemental abundances (for example). This may also potentially allow us to remove the need for the additional \lq\nh(scatt)\rq\ layer that was introduced in Section\,\ref{sec:modelT} for absorbing the scattered power law that dilutes EW(\fek). 

We first confirmed that simply decoupling the normalisation constant of the fluorescence line $A_L$ from the normalisation of the Compton scattering component $A_S$ allowed a very good fit (with a fitted sub-unity value of $A_L$\,=\,0.26\,\p\,0.06) with no extra PL$_{\rm scatt}$ component required, equivalent to model P. The problem is that such a decoupling cannot be self-consistently interpreted as having sub-Solar abundances, because of the assumption of Solar abundances for the Compton scattering component that produces the overall continuum shape of the Compton hump. 

We also attempted more complex scenarios, including the possibility of having multiple scatterers. Such a scenario includes two components inclined at orthogonal angles of 0 [face-on] and at 90\,deg [edge-on] with $A_S$\,=\,$A_L$ for each component, and is discussed at length by \citet{yaqoob12}. But these models appeared to require extreme decoupling between the two scatterers. For instance, allowing freely varying cross-normalisation constants between these scatterers, the face-on reflection component ($A_{S0}$ in Yaboob's terminology) prefers a normalisation $\gtsim$\,5\,times stronger than the edge-on component ($A_{S90}$) responsible for the l.o.s. obscuration. This implies a strong departure from the default time-steady illuminated \mytorus\ geometry with covering factor of 0.5. Such a scenario with strong variability on characteristic timescales relevant to the inner torus is discussed and argued against in Section\,\ref{sec:switchedoffscenario}.

\begin{figure*}
  \begin{center}
    \includegraphics[angle=270,width=11.5cm]{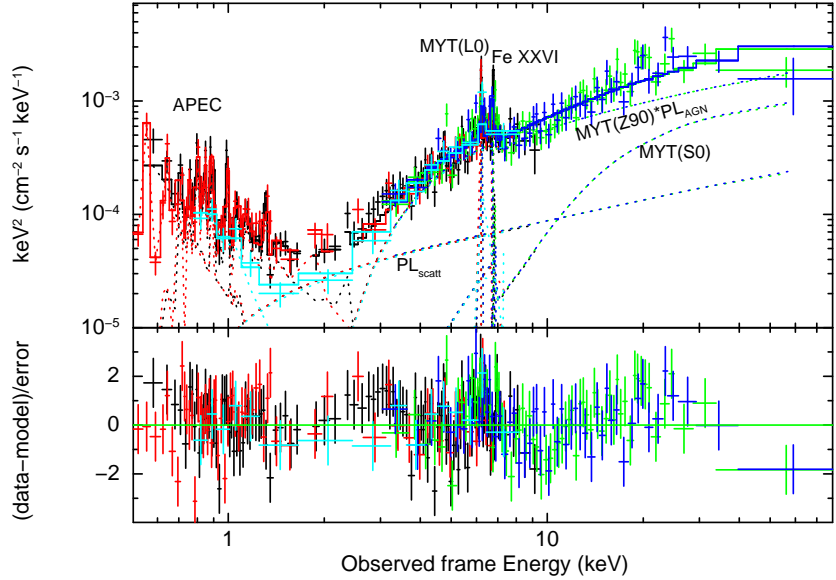}
\caption{Decoupled \mytorus\ model fit with the direct component absorber (MYTZ90) orientation being 90\,deg, and the reflection being dominated by the face-on scatterer MYTS0 and its corresponding fluorescence emission from MYTL0. 
 \label{fig:decoupled_likeyaqoob15}}
  \end{center}
\end{figure*}

Alternatively, the column densities between the two scatterers may also be untied. This is equivalent to simulating a clumpy medium with the line-of-sight obscuration differing from global. Such a solution is shown in Fig.\,\ref{fig:decoupled_likeyaqoob15}. The spectrum can be crudely fit ($\chi^2$/dof\,=\,371/301) with a low l.o.s column associated with the edge on obscurer \nh(nuc)\,=\,1.3\,(\p\,0.3)\,$\times$\,10$^{23}$\,cm$^{-2}$, This component is denoted by \lq MYT(Z90)\rq\ in the figure, as in \citet{yaqoob12}. However, this also requires a very hard photon index for PL$_{\rm AGN}$, with $\Gamma$\,=\,1.40$_{-u}^{+0.08}$ pegged at the lower limit allowed by the model. The out-of-sight global column from the face-on scatterer is \nh(S0)\,=\,2.4$_{-0.8}^{+1.7}$\,$\times$\,10$^{24}$\,cm$^{-2}$ (with corresponding Compton scattered and fluorescence components denoted by \lq MYT(S0)\rq\ and \lq MYT(L0)\rq; Ibid.). There is more than an order of magnitude difference between the two column densities. In order to prevent the strong normalisation departure mentioned in the preceding paragraph, we limited $A_{S0}$ to a maximum value of 1.20, and the fit does, indeed, want to exceed this limit, with $A_{S0}$ pegged at 1.20$_{-0.07}^{+u}$. The scattered fraction is high, with \fscatt\,=\,0.12$_{-0.06}^{+0.04}$ but we stress that there is no additional ad hoc screen associated with the aforementioned \nh(scatt) component in this decoupled model (or in any of the others discussed in this section). 
The overall fit statistic is much worse ($\Delta$\,$\chi^2$\,=\,+46 for a single extra dof) than the default model M solution shown in Table\,\ref{tab:x}. 

Scenarios with strong clumpiness have also been discussed for Mrk\,3 \citep{yaqoob15, guainazzi16}. But such a scenario is not obviously applicable to NGC\,7674. Mrk\,3 is known to be continuously variable in flux and line-of-sight column density, whereas NGC\,7674 has been stable since the \sax\ observations in 1996. Furthermore, it is clear from the discussion of the pegged photon index and normalisation values above that the fit is attempting to converge on a harder, more reflection-dominated continuum shape, suggesting that it does prefer a higher line-of-sight column density.

Unless the direct {PL}$_{\rm AGN}$ component can be {\em robustly} detected, such solutions are highly complex and degenerate, and do not yield any obvious physically useful insights on the nuclear medium beyond our three canonical models discussed in Table\,\ref{tab:x}. This is particularly true for reflection-dominated AGN with no direct detection of PL$_{\rm AGN}$  
(e.g., see detailed discussion in \citealt{yaqoob12} on this point). Therefore, although we cannot rule out more complex models at this stage, we do not investigate the decoupled modes in greater detail herein. Future high signal-to-noise broadband spectra, and robust detections of reflection features such as the fluorescence Compton shoulder with X-ray calorimeters, could yield more insight on such models. 

\begin{table*}
  \begin{center}
  \caption{Results of X-ray spectral fitting to NGC\,7674\label{tab:x}}
  \begin{tabular}{lccccr}
\hline
Component          &  Parameter       &     Model P         & Model T     &    Model M   & Units \\
\hline
\hline
Primary Absorber/Reflector & \nh(nuc) & $3.4_{-0.6}^{+0.8}$   &  $36_{-10}^{+u}$ & 5.5$_{-2.3}^{+u}$  &  10$^{24}$ cm$^{-2}$\\
                   & \nh(eq)          &    $''$             & $''$            & 10.0$_{-4.1}^{+u}$  &  10$^{24}$ cm$^{-2}$\\
                   & $\theta_{\rm inc}$ & 85$_{-8}^{+u}$        & $87.1^f$          &  $65_{-4}^{+9}$   & deg\\
                   & $\theta_{\rm tor}$ & --                  & 62$_{-11}^{+15}$  &  --            & deg\\
                   & $R$              & --1$^f$             & --              & --             &  \\
                   & $A_{\rm Fe}$       & 0.51$_{-0.11}^{+0.10}$ & --$^\ddag$              & --$^\ddag$             &  \\
                   & EW(Fe\,K$\alpha$)& \multicolumn{3}{c}{................ 0.38$_{-0.09}^{+0.10}$ ................$^\dag$}     &  keV \\
                   & EW(\fexxvi)& \multicolumn{3}{c}{................ 0.20$_{-0.09}^{+0.11}$ ................$^\dag$}     &  keV \\
{PL}$_{\rm AGN}$& $\Gamma$         & $2.07_{-0.11}^{+0.15}$ & 1.80$_{-0.11}^{+0.15}$ & 1.93$_{-0.12}^{+0.28}$ &  \\
\multicolumn{6}{c}{}\\
\multicolumn{6}{c}{\em Additional components}\\
Host absorption  & \nh(host)   & 4.0$_{-1.6}^{+2.1}$    & 3.1$_{-1.1}^{+1.7}$   &  3.4$_{-0.2}^{+0.2}$   & 10$^{21}$\,cm$^{-2}$ \\
\apec$_1^a$          & $kT_{\rm apec_1}$              & $0.11_{-0.02}^{+0.03}$ & 0.12$_{-0.03}^{+0.03}$ & $0.11_{-0.04}^{+0.04}$ & keV\\
                   &   $A_{\rm apec_1}$  & $0.1_{-0.08}^{+u}$     & 0.1$_{-0.09}^{+0.6}$  & $0.1_{-0.07}^{+0.8}$ & \\
\apec$_2^a$          & $kT_{\rm apec_2}$              & $0.56_{-0.07}^{+0.07}$ & 0.59$_{-0.07}^{+0.07}$ & $0.59_{-0.08}^{+0.09}$ & keV\\
                   &   $A_{\rm apec_2}$  & $0.3_{-0.1}^{+0.8}$    & 0.2$_{-0.1}^{+0.5}$  & $0.3_{-0.2}^{+0.5}$ & \\
\apec$_3^a$          & $kT_{\rm apec_3}$              & $0.17_{-0.04}^{+0.05}$ & 0.15$_{-0.04}^{+0.06}$ & $0.14_{-0.04}^{+0.06}$ & keV\\
                   &   $A_{\rm apec_3}$  & $0.01_{-u}^{+0.07}$    & 0.01$_{-u}^{+0.11}$ & $0.01_{-u}^{+0.16}$ & \\
{\sc PL}$_{\rm soft}$& $\Gamma_{\rm soft}$&   --                & 2.0$^f$ & 2.0$^f$ & \\
                   &  Norm             &   --                & 2.5$_{-1.5}^{+1.2}$ & 2.5$_{-1.5}^{+1.4}$ & 10$^{-5}$\,ph\,keV$^{-1}$\,cm$^{-2}$\,s$^{-1}$ \\
Scattering   & \fscatt           &  --                 & 3.5\,\p\,1.2 & $6.1_{-3.1}^{+3.6}$ &  10$^{-2}$\\
             & \nh(scatt)  &    --               & 9.4\,\p\,2.5        &  10.6$_{-2.3}^{+3.3}$  & 10$^{22}$\,cm$^{-2}$ \\
$C_{\rm XIS\ FI}^{\rm XIS\ BI}$ cross-calib & {\sc const}    & 1.00\,\p\,0.07     & 1.01\,\p\,0.07 &  $1.00_{-0.07}^{+0.07}$ & \\
$C_{\rm XIS\ FI}^{\rm FPMA}$ cross-calib    & {\sc const}   & $1.06_{-0.07}^{+0.08}$ & 1.11\,\p\,0.08 & $1.12_{-0.08}^{+0.09}$ & \\
$C_{\rm XIS\ FI}^{\rm FPMB}$ cross-calib    & {\sc const}   & $1.04_{-0.07}^{+0.09}$  & 1.10$_{-0.08}^{+0.05}$ & $1.11_{-0.09}^{+0.09}$ & \\
$C_{\rm XIS\ FI}^{\rm XRT}$ cross-calib    & {\sc const}    & $0.98_{-0.13}^{+0.13}$ & 1.01\,\p\,0.14 & $1.00_{-0.14}^{+0.15}$ & \\
                   &                  & & &  & \\
$\chi^2$/dof       &                  & 325/305             & 320/300 & 325/300 & \\
\hline
\end{tabular}~\par
Model P: \pexmon\ component fit \citep{pexmon}.\\
Model M: \mytorus\ coupled component fit \citep{mytorus}.\\
Model T: \torus\ model component fit \citep{brightmannandra11}.\\
$^u$unconstrained to within the model limits. \nh\ upper limits are 10$^{25}$\,cm$^{-2}$ and 10$^{26}$\,cm$^{-2}$ for \mytorus\ and \torus, respectively. \apec\ is defined between abundances of 0 to 5, relative to Solar.\\
$^f$fixed.\\
$^\dag$Equivalent width measured using a simple Gaussian atop a locally fitted powerlaw continuum.\\
$^a$\apec$_1$ and \apec$_2$ represent thermal components in \suzaku\ XIS, and \apec$_3$ is for \swift\ XRT.\\
$^\ddag$These models are defined for abundances fixed to Solar only.
\end{center}
\end{table*}

\begin{figure*}
  \begin{center}
    \includegraphics[angle=270,width=11.5cm]{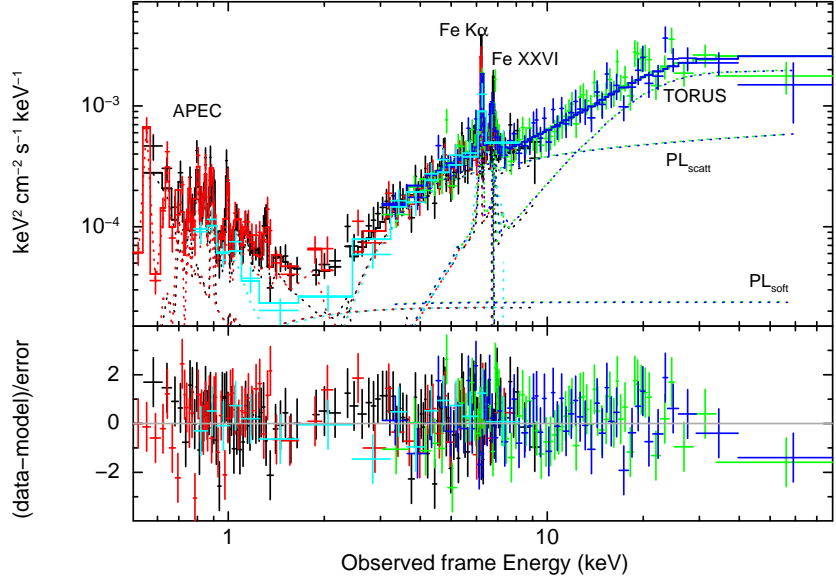}
\caption{\suzaku, \swift\ and \nustar\ data fits with the reflection model T. Colours are as in Fig.\,\ref{fig:pexmon}. 
 \label{fig:modelT}}
  \end{center}
\end{figure*}

\section{Discussion}




\subsection{Intrinsic luminosity}
\label{sec:luminosity}

Our three models return intrinsic $L_{2-10}$ values ranging over \,$\approx$\,(3--5)\,$\times$\,10$^{43}$\,erg\,s$^{-1}$ for the best fit parameters in Table\,\ref{tab:x}. The narrow range of these luminosities is noteworthy, despite the differing geometries inherent to these models. Placing this range in the context of other well-known CTAGN, the luminosity of NGC\,7674 is similar to NGC\,1068, and is factor of $\approx$\,2 lower than Mrk\,34 \citep{bauer15, g14}. The former object is the prototypical reflection-dominated AGN, while the latter is the most luminous known bona fide CTAGN within $\sim$\,250\,Mpc. For the {\em observed} (i.e. absorbed) spectra, we have $F_{2-10}^{\rm obs}$\,=\,7.7\,$\times$\,10$^{-13}$\,erg\,s$^{-1}$\,cm$^{-2}$, or $L_{2-10}^{\rm obs}$\,=\,1.5\,$\times$\,10$^{42}$\,erg\,s$^{-1}$ -- a factor of $\approx$\,20--30 times lower than the inferred intrinsic power. 

\begin{figure*}
  \begin{center}
    \includegraphics[angle=0,width=12cm]{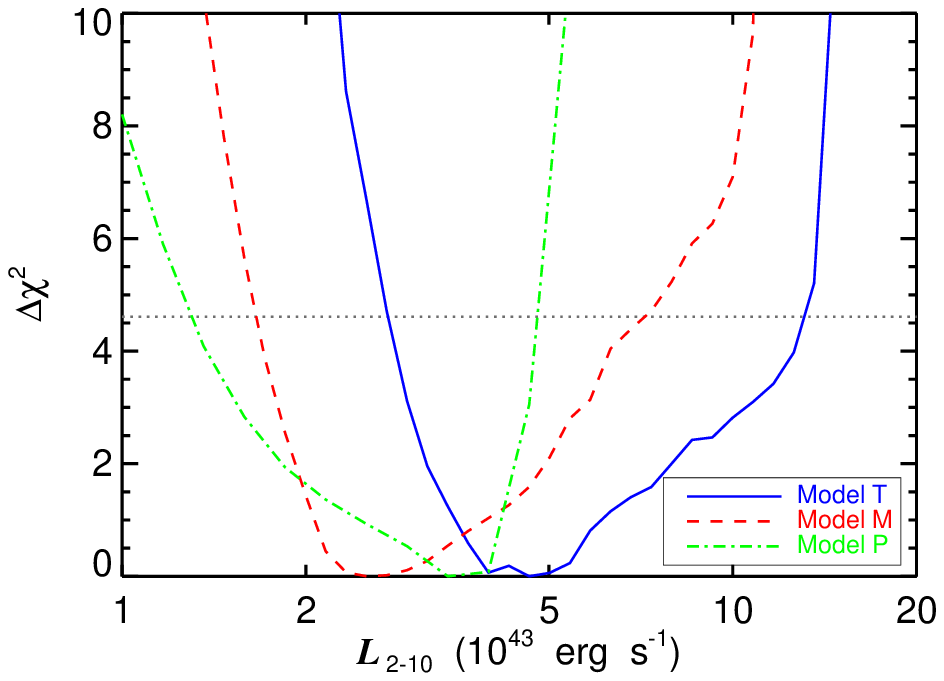}
\caption{$\chi^2$ contours as a function of $L_{2-10}$ for models T, M and P. Luminosities are computed over the the two-dimensional parameter space of randomised photon index $\Gamma$ and normalisation $N$ values for PL$_{\rm AGN}$, and the overall one-dimensional {\em envelope} of corresponding $\chi^2$ values is then plotted. The dotted horizontal line corresponds to $\Delta \chi^2$\,=\,4.61, or 90\%\ confidence for the two starting parameters $\Gamma$ and $N$, relative to the minimum $\chi^2$ for each model.
 \label{fig:lumchi2}}
  \end{center}
\end{figure*}

In order to estimate realistic uncertainties on the intrinsic luminosities, we stepped over a 2-dimensional grid of $\Gamma$ and normalisation ($N$) for the intrinsic {PL}$_{\rm AGN}$ -- the two parameters which determine the absorption-corrected flux (and hence direct luminosity). Carrying out fits over the grid yields a $\chi^2$ value for each combination of $\Gamma$ and $N$, and thus effectively for each value of $L_{2-10}$. Different combinations of the two starting parameters can return identical $L_{2-10}$ values, so the 1-dimensional space of $\chi^2$ as a function of $L_{2-10}$ is not unique. But the overall uncertainties can be gauged from the {\em envelope} of $\chi^2$ contours for all combinations. This envelope is plotted in Fig.\,\ref{fig:lumchi2} for both models M and T. The figure shows that realistic $L_{2-10}$ uncertainties span the range of $\approx$\,(1.3--13)\,$\times$\,10$^{43}$\,erg\,s$^{-1}$, or about an order of magnitude. 


Our estimate of the mass of the supermassive black hole (SMBH) in NGC\,7674 is \Mbh\,$<$\,10$^{7.43}$\,\Msun\ (see Appendix). Using our best-fit \ltwoten\ range of (3--5)\,$\times$\,10$^{43}$\,erg\,s$^{-1}$ together with a bolometric correction likely range of \lbol/\ltwoten\,$\approx$\,10--20 \citep{vasudevanfabian07}, we estimate an Eddington ratio range of \lbol/\ledd\,$>$\,0.09--0.29. Including the full range of X-ray luminosity uncertainties expands this range to \lbol/\ledd\,$>$\,0.04--0.74, implying the presence of an efficiently accreting AGN. 

\subsubsection{Multiwavelength comparisons}
\label{sec:correlations}

Multiwavelength scaling relations are very useful for comparison of intrinsic luminosity estimates, especially in the heavily Compton-thick regime where the direct X-ray emission is not detected. Two commonly used multiwavelength indicators of intrinsic AGN power are the mid-infrared continuum and the optical forbidden emission line luminosities, in particular the \oiii\,\l5007\AA\ line doublet \cite[e.g. ][ and references therein]{g09_mirxray, lamastra09}. We compare \obj\ to these relations here. 

NGC\,7674 has been observed at high angular resolution in the mid-infrared using 8\,m class telescopes. It is found to have a 12\,\micron\ infrared luminosity of $L_{12}$\,=\,1.8(\p 0.3)\,$\times$\,10$^{44}$\,erg\,s$^{-1}$ \citep{asmus14}. Using this value of $L_{12}$, the relation between infrared and X-ray luminosities 
for local AGN \citep{asmus15, g09_mirxray} predicts an intrinsic $L_{2-10}$ range of (6--9)\,$\times$\,10$^{43}$\,erg\,s$^{-1}$ at 68\,\% confidence, which is marginally higher than the mean best fit luminosities from our spectral analysis, but entirely consistent with the full confidence regions for both models shown in Fig.\,\ref{fig:lumchi2}. Assuming a 6\,\micron\ luminosity lower by a factor of 2 as compared to $L_{12}$ \citep[cf. ][]{goulding12} together with the $L_6$/\ltwoten\ relation by \citet{stern15} relevant for high luminosity AGN, the predicted \ltwoten\ decreases by a further 0.1\,dex. 
The angular resolution of the mid-infrared observations used (seeing-limited at $\approx$\,0.4\,arcsec) 
corresponds to a physical scale of $\approx$\,0.24\,kpc for the unresolved nuclear emission at the distance of NGC\,7674, and represents the best direct measure of the intrinsic AGN power, with emission from surrounding star formation being largely resolved out. A further check on any contamination by non-AGN components may be obtained from the mid-infrared colour of \obj\ as tabulated in the \wise/AllWISE catalogue \citep{wise, allwise}. The $W1$--$W2$ colour is 1.16\,\p\,0.03\,mag, which places the source above the colour threshold of $W1$\,--\,$W2$\,$>$\,0.8 identified by \citet{stern12} where the mid-infrared emission is likely to be AGN-dominated. 

Comparing next to the \oiii\ emission line, \citet{bassani99} present the \oiii\l5007 luminosity of the source as \loiii\,=\,3.5\,(\p\,0.2)\,$\times$\,10$^{42}$\,erg\,s$^{-1}$. This value is derived after correcting for dust reddening based upon a Balmer decrement of 4.80. 
Using the \ltwoten/\loiii\ relationship presented in \citet{lamastra09} with a scatter of 0.6\,dex, we expect $L_{2-10}$\,=\,5$_{-4}^{+7}$\,$\times$\,10$^{43}$\,erg\,s$^{-1}$ at 68\,\% confidence, overlapping well with our spectrally modelled intrinsic luminosity range. 

In summary, 
the best fit X-ray luminosity measured from our spectral analysis agrees 
well with the multiwavelength comparisons above, especially when considering the full range of realistic uncertainties on the spectral modelling (Fig.\,\ref{fig:lumchi2}). This is encouraging given that this source appears to be heavily Compton-thick and that the true torus geometry is unknown. 
Broadband spectral modelling of high signal-to-noise X-ray data is what enables us to get such good agreement. 


\subsubsection{The soft X-ray component}
\label{sec:soft}

  The absorption-corrected luminosity in the XIS 0.5--2\,keV band is $L_{0.5-2}^{\rm APEC}$\,$\approx$\,4.5\,$\times$\,10$^{42}$\,erg\,s$^{-1}$. This is for both \apec\ components combined, but is dominated by a factor of 4 by the lower temperature component. Using the relation for star-forming galaxies by \citet{mineo12_hotgas} between the thermal component luminosity and the star formation rate (SFR), we estimate an X-ray derived SFR$_{\rm X-ray}$\,$\approx$\,9\,$\times$\,10$^3$\,\Msun\,yr$^{-1}$.

  This may be compared to the infrared derived SFR, based upon the far-infrared continuum luminosity $L_{\rm IR}$ and its relation to SFR$_{\rm IR}$ \citep{kennicutt98}. For \obj, $L_{\rm IR}$\,$\approx$\,10$^{11.56}$\,\Lsun\ \citep{koss13}, which yields SFR$_{\rm IR}$\,$\approx$\,60\,\Msun\,yr$^{-1}$. This is more than two orders of magnitude lower than the estimated SFR$_{\rm X-ray}$, and implies that starburst-powered \apec\ components alone are an unphysical representation of the soft X-ray emission in NGC\,7674. 
  Photoionisation could instead power some of this, as we have already alluded to on several occasions. 
We also note that uncertainties related to absorption correction of soft X-rays cannot account for the extremely high SFR$_{\rm X-ray}$. Ignoring absorption corrections and using the observed (i.e. absorbed) $L_{0.5-2}^{\rm APEC}$ directly still results in SFR$_{\rm X-ray}$\,$\approx$\,950\,\Msun\,yr$^{-1}$, far higher than SFR$_{\rm IR}$.
  
We note, however, that the estimated value of SFR$_{\rm IR}$ is itself large. For example, it is about 10\,times above the SFR of the prototypical starburst galaxy M82 \citep[e.g. ][]{telesco80}. Such high star formation is likely to power extended ionised gas emission, and we will return to possible implications of this in Section\,\ref{sec:ionised}.

\subsection{The nature of the long-term flux changes}
\label{sec:lc}

Fig.\,\ref{fig:lc} shows the long-term X-ray light curve of \obj\ over a period of about 37\,years. As first noted by \citet{bianchi05}, the source showed a decline by a factor of $\sim$\,3 in the 2--10\,keV band between the first detection by \heao\ in the late 1970s and the \ginga\ measurement in 1989, followed by a further order of magnitude flux decrease in 1996 when \sax\ identified the source as a CTAGN \citep{malaguti98}. Thereafter, the source flux has remained constant with no significant spectral or flux variation for the past $\approx$\,20 years. This now includes the most recent \suzaku\ and \nustar\ observations. The individual \swift\ observations spanning the period of 2011--2014 detailed in the Appendix also show fluxes or detection limits broadly consistent with \sax, \suzaku\ and \nustar. Finally, the custom analysis of the \swift/BAT maps by \citet{koss13} shows a weak detection with flux $F_{14-195}$\,=\,9.9$_{-2.4}^{+5.1}$\,$\times$\,10$^{-12}$\,erg\,s$^{-1}$\,cm$^{-2}$ consistent with \nustar. For instance, our model T, when extrapolated to the energy range of 14--195\,keV, implies best fit fluxes ranging over (8.4--9.3)\,$\times$\,10$^{-12}$ erg\,s$^{-1}$\,cm$^{-2}$ between the mission cross-calibration uncertainties. 

Here, we examine the viability of the inferred long-term decline and its implications.

\subsubsection{On the possibility of contamination by another source}

Could the pre-\sax\ flux decline be associated with a contaminating source, unrelated to the AGN? The observed luminosities at the \heao\ and \ginga\ epochs are above 10$^{42}$\,erg\,s$^{-1}$ -- already too high to be easily associated with X-ray binaries (XRBs) within the host galaxy. An XRB within our Galaxy which just happens to lie along the same l.o.s. as \obj\ cannot be ruled out, though the high Galactic latitude ($b$\,=\,--48$^{\circ}$) makes this unlikely. 

Regarding contamination by other AGN, 
there is only one possible bright source in the recent 70\,month all sky \swift/BAT hard X-ray survey that could potentially have contaminated both the \heao\ and \ginga\ fields of view. This is PKS\,2325+093 which lies at a separation of 0.9\,deg from \obj\ and shows a BAT flux of $F_{14-195}$\,$\approx$\,3\,$\times$\,10$^{-11}$\,erg\,s$^{-1}$\,cm$^{-2}$ \citep[][]{baumgartner13}. 
However, this source lies outside the \heao\ positional error box of \obj\ \citep{bianchi05}. Moreover, PKS\,2325+093 shows a very hard spectrum, at least in the BAT band, with $\Gamma_{\rm BAT}$\,=\,1.29\,\p\,0.28. This is much harder than the spectral shape inferred for the \ginga\ observation below $\approx$\,10\,keV by \citet{bianchi05}. Although a drastic change in spectral curvature around 10\,keV cannot be ruled out, the combined weight of evidence appears to disfavour contamination. 

\subsubsection{Background uncertainties}

Measurements with non-imaging detectors such as the \heao\ A--1 Large Area Sky Survey instrument \citep{heaoa1} and the \ginga\ Large Area Counter \citep{gingalac} are prone to uncertain background estimates, especially at low source flux levels. However, as discussed by \citet{bianchi05}, the background for the observation of \obj\ is based on a scanning observation obtained close in time to the target \citep{awaki91}. These background scans result in noise estimates which are more reliable than model estimates typically adopted for non-imaging detectors. 

Therefore, the \ginga\ flux, at least, is considered to be reliable and is significantly higher than that seen by subsequent missions, by about a factor of 10. The \heao\ flux, on the other hand, may well be affected by the above uncertainties, and we do not consider it as a strong constraint in the following discussion.

\subsubsection{The \lq switched-off\rq\ AGN scenario}
\label{sec:switchedoffscenario}

Assuming that the luminosity change (at least that between \ginga\ and subsequent missions) is associated with the AGN in \obj\ itself, there are then two possible implications: the AGN could either have faded dramatically (an effective \lq switch-off\rq, as is referred to hereafter), or it could have become enshrouded within Compton-thick material after the \ginga\ observation. We consider these cases here. 

The switched-off AGN scenario was examined by \citet{bianchi05}, motivated by the fact that, unlike other changing-look AGN in which \nh\ variations occur relatively frequently, \obj\ did not show any flux or spectral variations in post--\sax\ observations. In this scenario, the observed reflection component dominating at hard X-rays is expected to be delayed with respect to the incident {PL}$_{\rm AGN}$ which has now faded. Similarly, the absorbed scattered component (PL$_{\rm scatt}$)\rq\ included in our models T and M (cf.\,Fig.\,\ref{fig:modelT}) would also be interpreted as a delayed component scattered in to the l.o.s from material on $\sim$\,pc scales, or larger. The current {PL}$_{\rm AGN}$ flux level cannot be higher than that this PL$_{\rm scatt}$ component, which has a deabsorbed luminosity $L_{2-10}^{\rm scatt}$\,$\approx$\,2\,$\times$\,10$^{42}$\,erg\,s$^{-1}$ -- an order of magnitude fainter than inferred during the \ginga\ observation epoch (1989). 
In model P, where no scattering component is required (cf.\,Fig.\,\ref{fig:pexmon}), the constraint on the current intrinsic AGN luminosity is even more stringent, with $L_{2-10}$ expected to be 10\,times lower still. 

Assuming that the source switched off between the \ginga\ (1989) and \sax\ (1996) epochs around 1993, i.e. 21\,\p\,3 years before the \nustar\ observation, the absence of any flux change since then places a minimum limit on the radius ($R$) of an axisymmetric reflector $R$\,=\,3.2\,pc based upon a simple consideration of the light travel time from the far wall of a nearly edge-on reflector. 
Considering the average travel time over the full body of the reflector, and/or intermediate inclinations angles, would push up the lower limit on $R$. While extended reflectors on scales of up to $\sim$\,150\,pc have been observed in several CTAGN (e.g. NGC\,4945, \citealt{marinucci12}; Circinus, \citealt{arevalo14}; NGC\,1068, \citealt{bauer15}), the studies so far find that these extended components make relatively minor flux contributions compared to the compact reflectors.

In fact, detailed studies of Type 1 AGN have shown that the bulk of the neutral \fek\ emission line arises at very compact scales of the dust sublimation radius (\rsub) or smaller \citep[see detailed discussion in ][ and references therein]{g15_fek}, and in the orientation-based unification scheme, this would also hold for obscured and CTAGN viewed at higher inclination angles. For \obj, \rsub\ is estimated to be $\approx$\,0.1$_{-0.04}^{+0.05}$\,pc based upon infrared luminosity scaling relations determined at high angular resolution where the AGN can be effectively isolated from surrounding star formation \citep{hoenig10}. This size scale is about 30\,times smaller than the lower limit on $R$ above. 

\begin{figure*}
  \begin{center}
    \includegraphics[angle=0,width=12cm]{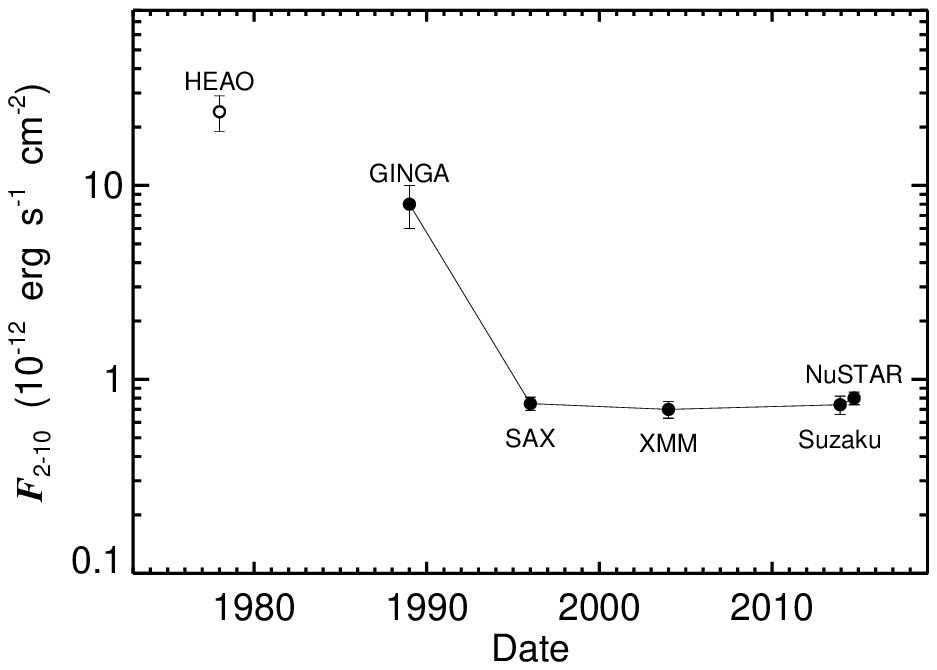}
\caption{Long-term 2--10\,keV observed flux light curve of \obj\ spanning $\sim$1977/78 (the \heao\ epoch) to late 2014 (\nustar). The \heao\ measurement is segregated from the other fluxes to stress its potentially uncertain nature. 
 \label{fig:lc}}
  \end{center}
\end{figure*}

In other words, in the switched-off AGN scenario, a typical (sub)-pc scale torus reflector would be expected to respond to a decline of the intrinsic continuum on timescales faster than seen in Fig.\,\ref{fig:lc}, and we would have expected to see some change in the reflected component fluxes under the switched-off AGN scenario by now. 




\subsubsection{Clumpy Compton-thick obscurer scenario}
\label{sec:obscuringscenario}
  
In the obscured AGN scenario, the source became obscured sometime between the \ginga\ (1989) and the \sax\ (1996) epochs, and is now fully covered by CT material along the l.o.s. Varying nuclear obscuration is common in AGN, but extreme transitions between Compton-thick and Compton-thin states are not \citep{markowitz14}. So it is interesting to examine whether the obscurer in NGC\,7674 may somehow be atypical. Some qualitative constraints on the nature and geometry of the obscuring clouds are possible as follows. 

Firstly, we argue that the global covering factor of the material is unlikely to be atypically low. This is because the reflection component we observe is strong. The observed reflected luminosity ($L_{\rm obs}$) scales approximately with intrinsic luminosity ($L_{\rm int}$) as 

\begin{equation}
L_{\rm obs} \sim L_{\rm int}\ \Omega\  a
\end{equation}

\noindent
where $\Omega$ is the solid angle of the reflector and $a$ is the albedo. The physical models that we have used in our spectral fits assume fairly typical geometrical covering factors for the torus, and yield estimates of $L_{\rm int}$ which are in reasonable agreement (to within a factor of a few) with other multiwavelength indicators (\S\,\ref{sec:correlations}). If the reflector had small covering factor $\Omega$, the corresponding geometrical correction would imply much higher values of $L_{\rm int}$. 

The complete lack of recent X-ray flux variability also supports this. Whereas an extended distribution of clouds can naturally spread out and dampen variations in the reflected flux relative to any variability in the direct AGN radiation, this is not possible in the low covering factor limit. Indeed, most reflection-dominated CTAGN are observed to show little, or no flux variations at all (cf. two well-studied examples include Circinus [\citealt{arevalo14}] and NGC\,5643 [\citealt{annuar15}], although sensitive observations have recently caught a transient column density change in NGC\,1068 [\citealt{marinucci16}]). Sources where significant flux variability is observed (e.g. NGC\,4945 [\citealt{puccetti14}], ESO\,565--G019 [\citealt{g13_eso565}], and IC\,751 [\citealt{ricci16_ic751}]) tend to be {\em mildly} Compton-thick AGN or significantly clumpy, with the direct transmitted component being stronger than the reflection component over at least some portion of the hard X-ray regime. In particular, many observations support the presence of a small $\Omega$ for the obscurer in NGC\,4945 \citep[e.g. ][ though there are other possible scenarios as discussed by \citealt{brightman15}]{madejski00}.


The emerging scenario for NGC\,7674 then is that a patchy distribution of clouds obscures the nucleus with a covering factor fairly typical of standard torus models. By the time of the \sax\ observation in 1996, a cloud ensemble with CT column density had fully obscured the l.o.s to the nucleus, and this ensemble has continuously covered the l.o.s. ever since.


Unlike changing-look AGN, however, \obj\ does not show frequent flux or spectral variability. This raises a complementary point of view that the past high state caught by \ginga\ was instead a transient near-complete {\em unveiling} of the nucleus caused by the l.o.s passing through a hole in the patchy obscurer. Assuming relatively sharp optical depth edges to the obscuring clouds, the time gap of $\approx$\,7\,years between the \ginga\ and \sax\ observations is probably a strong upper limit to the time taken to cover the AGN X-ray emission region, which is expected to be highly compact \citep[e.g. ][]{risaliti07}. In other words, the \lq unveiling\rq\ could have been very brief, and {\em Ginga} may have simply been fortunate to catch the event.

Interestingly, converting the observed 1989 \ginga\ flux of 8(\p2)\,$\times$\,10$^{-12}$\,erg\,s$^{-1}$\,cm$^{-2}$ \citep{bianchi05} to a luminosity yields $L_{2-10}$\,$\approx$\,2\,$\times$\,10$^{43}$\,erg\,s$^{-1}$, again consistent with our spectral analysis results based upon the more recent data (Section\,4.1). The effect of absorption in the \ginga\ spectrum appears minimal. This consistency provides additional support for the obscured AGN scenario. 

Finally we note that, given the uncertainty associated with the \heao\ flux measurement, there is no strong constraint on the pre-\ginga\ flux evolution. If the \heao\ flux measurement were to be correct, the above analysis would imply that the AGN would have been {\em intrinsically} more luminous by a factor of $\approx$\,3 during the \heao\ era. In this case, explaining the long-term flux evolution would require a combination of source fading (between \heao\ and \ginga) followed by patchy obscuration, which seems unlikely.

\subsection{On the weakness of the \fek\ line}

In our analysis, we found that the source has a relatively weak neutral \fek\ emission line, and showed that this can be reproduced by assuming either a low elemental abundance (model P), or an absorbed scattering solution (models T and M). Here, we first place this result in context of other local CTAGN (Section\,\ref{sec:ewhist}) and then briefly investigate one other potential scenario of Iron line dilution by a jet (Section\,\ref{sec:jet}). In Section\,\ref{sec:ionised}, we extend the comparison with other objects to higher redshifts (especially those that show prominent {\em ionised} Fe lines), before finally examining the broader implications of Fe line measurements in deep AGN surveys in Section\,\ref{sec:implications}. 

\subsubsection{Comparison with other {\em bona fide} CTAGN}
\label{sec:ewhist}

Compton-thick AGN are usually associated with strong \fek\ emission lines with equivalent widths (EW) of $\gtsim$\,1\,keV. Our work shows a much weaker EW\,$\approx$\,0.4\,keV \citep[also reported by ][]{bianchi05}, but our spectral modelling of \obj\ also consistently finds high Compton-thick column densities using several different spectral models. 

\begin{figure}
  \begin{center}
    \includegraphics[angle=0,width=8.5cm]{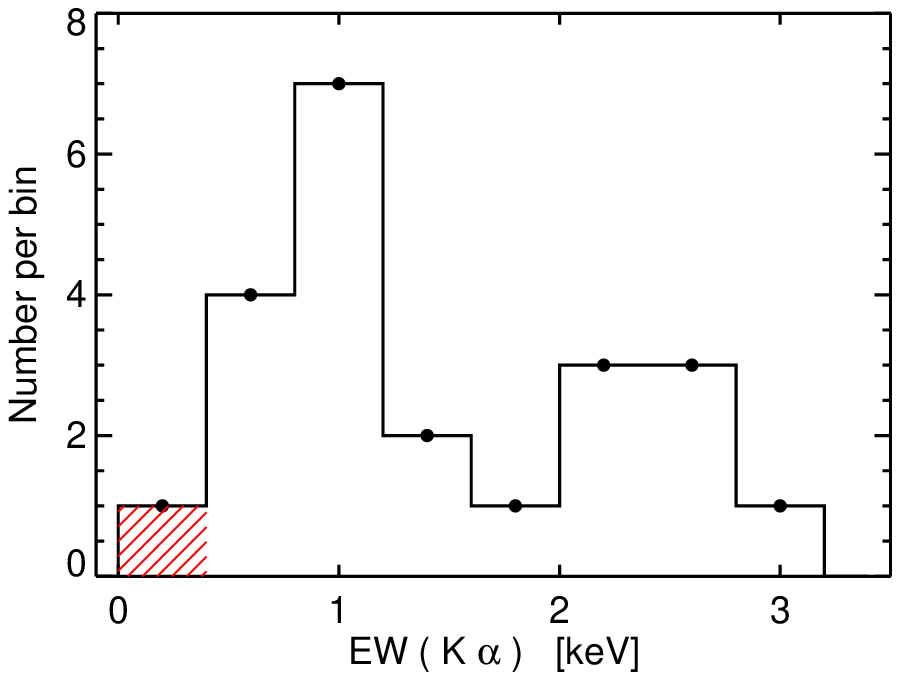}
\caption{Distribution of equivalent widths (EWs) of the neutral \fek\ emission line for the bona fide CTAGN compiled in \citet{g14}. The red hatched region corresponds to NGC\,7674 (EW\,=\,0.38$_{-0.09}^{+0.10}$\,keV). 
 \label{fig:ewhist}}
  \end{center}
\end{figure}

In fact, comparing to other reflection-dominated AGN (such as NGC\,1068, NGC\,5643, Mrk\,34 and others with \nh\,$\gtsim$\,10$^{25}$\,cm$^{-2}$) amongst the local {\em bona fide} CTAGN population \citep{dellaceca08, goulding12, g14}, NGC\,7674 appears to have one of the weakest neutral Fe fluorescence lines. This is demonstrated in Fig.\,\ref{fig:ewhist}, where NGC\,7674 stands out in the distribution of EW distribution of bona fide CTAGN compiled by \citet{g14}. The EW values are taken from relevant recent references in the same paper, or (where recent values are not published) from the compilation of \citet{dellaceca08}. They are mostly based upon power law fits to the continuum together with Gaussian line components. Three sources have been excluded because of published works questioning their bona fide CTAGN nature, or showing that they have highly complex geometries: Mrk\,3 (EW\,$\approx$\,1.0\,\p\,0.3\,keV; \citealt{yaqoob15, guainazzi16}), NGC\,7582 (EW\,$\approx$\,0.6$_{-0.1}^{+0.6}$\,keV; \citealt{rivers15_7582}); and NGC 4939 (EW\,$\approx$\,0.5$_{-0.2}^{+0.4}$\,keV; \citealt{maiolino98}). Their exclusion does not affect the relative position of NGC\,7674.

\subsubsection{Dilution by a jet?} 
\label{sec:jet}

One other possibility may be that a jet is diluting the continuum underlying the \fek\ emission line and hence effectively weakening the observed EW. Such dilution has been inferred in several broad-line radio galaxies \citep[e.g. ][]{eracleous00}, and also in broad absorption line quasars \citep[e.g. ][]{luo13}. 

In the case of NGC\,7674, however, we consider this possibility to be unlikely for several reasons. Although \obj\ is a known radio source, it is classified as radio quiet \citep{xu99}. 
Then there is the distinct lack of X-ray variability at 2--10 keV as probed by several missions over the past $\sim$\,20\,years. 
Non-thermal emission from a jet is expected to be significantly variable, especially when probed on multiple (long) timescales that we are now able to sample (Fig.\,\ref{fig:lc}). 

Finally, one may try to quantify the expected flux from a jet irrespective of the above considerations. 
The core of NGC\,7674 has a morphology extended over $\approx$\,1\,kpc, characteristic of radio galaxies \citep[e.g. ][]{momjian03}. There is evidence of complex interactions of the radio ejecta with the interstellar medium -- complex enough that the location of the AGN itself is unclear in the radio \citep{momjian03}. The reported integrated 5\,GHz nuclear radio flux density is 67\,mJy \citep{condon91_ugc}, corresponding to a monochromatic luminosity density of $L_{\rm 5\,GHz}$\,=\,1.3\,$\times$\,10$^{30}$\,erg\,s$^{-1}$\,Hz$^{-1}$. 

The above luminosity is lower than the power typically associated with radio-loud (RLQs) as well as radio-intermediate quasars (RIQs). For example, \citet{miller11} studied samples of RIQs and RLQs and showed that there exists an \lq X-ray excess\rq\ above that seen in radio-quiet sources. The excess scales with radio luminosity and could be associated with a jet contribution to X-rays. At the observed luminosity density of NGC\,7674, the correlation between the X-ray excess and radio luminosity found by \citet{miller11} is consistent with no jet-related X-ray excess. Their correlation needs to be extended below their lower luminosity limit and would instead predict an X-ray deficit of --0.14\,\p\,0.09 in their adopted units of $\ell_x$\,--\,$\ell_{x,{\rm RQQ}}$. 

Instead of using the correlation for quasars, one may instead try using the correlation between radio power and the jet-related unabsorbed X-ray powerlaw luminosities for lower luminosity 3CRR sources presented by \citet[][]{hardcastle09}. In the terminology used by \citeauthor{hardcastle09}, the X-ray power law luminosity is $L_{\rm X,u}$, with the X-ray band being $\approx$\,0.4--8\,keV, and a fixed spectral slope $\Gamma_{\rm jet}$\,=\,2. From the reported 5\,GHz radio flux density, 
we predict $L_{\rm X,u}$\,$\approx$\,10$^{41}$\,erg\,s$^{-1}$, or $F_{2-10}$\,=\,3\,$\times$\,10$^{-14}$\,erg\,s$^{-1}$\,cm$^{-2}$. This is more than a factor of 20 fainter than our observed $F_{2-10}$. Accounting for a scatter of 0.56\,dex in the relation of \citet{hardcastle09}, the predicted jet X-ray flux is still lower than the observed flux by a factor of 6, rendering this scenario unlikely.

\subsubsection{On the strength of ionised Fe and the relation to other powerful infrared galaxies}
\label{sec:ionised}

A variety of studies have found that distant sources with high bolometric power, including ultraluminous infrared galaxies (ULIRGs) and sub-mm galaxies, show preferentially strong ionised Fe lines such as \fexxv\ (6.7\,keV) and \fexxvi\ (6.97\,keV; e.g. \citealt{iwasawa09, lindner12, gilli14}). The neutral \fek\ (6.4\,keV) line is extremely weak or undetected in many of these systems. This may be attributable either to the neutral reflection component being heavily embedded within the torus as seen by us, or to a highly ionised interstellar medium that is being energised by starburst activity, or a combination of the two effects. \citet{iwasawa12} further suggest that preferentially strong ionised emission lines may be connected to high accretion rates on the central AGN. 

With a bolometric infrared power of 10$^{11.56}$\,\Lsun\ \citep{koss13}, \obj\ lies in the regime of luminous infrared galaxies. This is lower than, but approaching, the regime associated with ULIRG luminosities. Such an intermediate luminosity could explain why the source displays {\em both} a neutral and an ionised Fe line with strengths comparable to within a factor of about 2. Our estimate of the Eddington ratio of the AGN in \S\,\ref{sec:luminosity} is as high as $\approx$\,0.4, and this could be even higher if \Mbh\ has been overestimated (see Appendix). \citet{brightman16} also compile evidence showing that some CTAGN (with steep X-ray power law photon indices) may exhibit high Eddington fractions. If so, \obj\ would be a local analogue of the more bolometrically luminous systems with high accretion rates studied by \citet{iwasawa12}. A more robust \Mbh\ estimate will be needed to test these parallels. 

  Finally, we note that a strong ionised \fexxvi\ emission feature may be accompanied by an ionised underlying continuum. So it is possible that the scattered component introduced in Section\,\ref{sec:modelT} actually arises from from photoionised or hot collisionally ionised gas which also self-consistently produces \fexxvi\ and some of the soft emission. The possible need for a photoionised component was discussed in Section\,\ref{sec:soft}. We attempted to fit the data as such, by replacing the absorbed PL$_{\rm scatt}$ component from our base model T with a photoionised model based upon the publicly-available CLOUDY code \citep{cloudy0507}. 
Examples of such fits can be found for various other AGN, including Mrk\,573 \citep{bianchi10} and ESO\,138--G001 \citep{decicco15}. Reasonable fits were possible without PL$_{\rm scatt}$, but we found that the photoionised component was required to be extremely strong relative to the Compton-thick torus, and that it dominated over the entire energy range to up $\sim$\,10\,keV. Although we cannot rule out such a model, photoionised components are usually much fainter \citep{bianchi05, decicco15}. High spectral resolution observations will be required to test this scenario in further detail.

\subsection{Implications for identifying Compton-thick AGN at low X-ray signal-to-noise}
\label{sec:implications}


The fact that NGC\,7674 shows a weak \fek\ emission line, yet clearly prefers a Compton-thick l.o.s column, has potentially important implications 
for the study of more distant AGN in deep surveys where the signal-to-noise is typically much weaker than in our data. Using simulations, \citet{koss15} found that robust characterisation of typical nearby CTAGN such as NGC\,3393 is only possible at relatively low redshifts of $z$\,$\ltsim$\,0.2 if one relies upon detection of the \fek\ emission line and broadband continuum for spectral modelling.

Our results on NGC\,7674 further complicate this issue. With a value of EW(\fek) weaker by a factor of $\approx$\,3 than in NGC\,3393, it becomes even more difficult to identify an object as being Compton-thick. We demonstrate this by carrying out a simulation of a source with a model spectrum identical to \obj, but with a lower flux of $F_{8-24}$\,=\,5\,$\times$\,10$^{-13}$\,erg\,s$^{-1}$\,cm$^{-2}$. Although four times fainter than \obj\ in the same band, this flux level lies more than an order of magnitude above the deepest flux level being probed in ongoing deep and wide \nustar\ surveys \citep{mullaney15, civano15, aird15, harrison16}. We simulated \nustar\ spectra for both FPMs using the default background and response simulation files\footnote{{\tt http://sc.nustar.caltech.edu}} provided by the \nustar\ team, and assuming an exposure time of 100\,ks. We also introduced a small redshift $z$\,=\,0.25, effectively assuming an intrinsic X-ray luminosity \ltwoten\,$\approx$\,10$^{45}$\,erg\,s$^{-1}$, characteristic of luminous AGN likely to be found in deep surveys. 

The simulated spectra are shown in Fig.\,\ref{fig:sim}. The rising continuum slope (from the Compton hump of our baseline model T) is clearly visible. We first fitted the simulated data with a simple power law model (not shown), which yielded $\Gamma$\,=\,1.1\,\p\,0.2 with $\chi^2$/dof\,=\,27/29. No additional \fek\ emission line is required for the fit. This hard slope is suggestive of the need for obscuration. However, including redshifted photoelectric absorption and reflection with a {\sc torus} model implies an \nh\ of only 2.2$_{-0.8}^{+0.9}$\,$\times$\,10$^{23}$\,cm$^{-2}$ for a fixed canonical $\Gamma$\,=\,1.8, i.e. a Compton-thin solution (the solution shown in the figure). 
This is a direct result of the lack of a strong \fek\ emission line. 

\begin{figure}
  \begin{center}
    \includegraphics[angle=270,width=8.5cm]{f8.ps}
\caption{\nustar\ simulated spectra for two FPMs assuming our model T best fitting model to NGC\,7674, but with a flux lower by a factor of 4 and shifted to $z$\,=\,0.25. The simulation is for 100\,ks of exposure time per FPM. The histograms show best fitting \torus\ model to the faked data, and are fitted with \nh\,=\,2.2$_{-0.8}^{+0.9}$\,$\times$\,10$^{23}$\,cm$^{-2}$ and $\Gamma$\,=\,1.8 for canonical fixed values of \thetator\,=\,60\,deg and \thetainc\,=\,87\,deg, i.e. a Compton-thin solution. The fit statistic is $\chi^2$/dof\,=\,34/29.
 \label{fig:sim}}
  \end{center}
\end{figure}

The newer torus models are now being widely used by the community for self-consistent modelling of the nuclear obscuring material, and these have proven to be very successful at characterising objects well into the Compton-thick regime. However, most of the publicly available torus models do not allow varying elemental abundances which, as we discuss in \S\,\ref{sec:modelT}, means that sources like NGC\,7674 cannot be fit without additional complexity. In any case, such model complexity is often not viable for fitting low signal-to-noise data, with the result that the column density, and hence intrinsic luminosity, of distant AGN may be underestimated. 

A full assessment of the resultant bias in deep surveys is beyond the scope of the present paper, until the frequency of weak \fek\ CTAGN such as NGC\,7674 can be established. Independent selection of large samples of CTAGN candidates based upon X-ray spectral diagnostics and on multiwavelength indicators (i.e. high ratios of \lmir\ or \loiii\ to observed \ltwoten) could be an informative first step in this direction (cf. \citealt{rovilos14}). Similarly, we recommend the use of varying elemental abundances as an additional free parameter in model fitting when signal-to-noise allows. 

If we are to make robust progress on the Compton-thick selection problem, multiple redundant cross-checking methods must be employed (the work of \citealt{brandt15} contains a recent, comprehensive analysis of various techniques in the literature). 

\section{Summary}

We have presented \nustar\ spectroscopy of the local reflection-dominated AGN NGC\,7674. Together with unpublished \suzaku\ and \swift\ data, we carried out broadband X-ray modelling of the 0.5--78\,keV spectrum assuming three geometries of the nuclear obscurer/reflector. The best fitting model in all cases requires a nuclear column density of obscuring gas \nh(nuc) of at least 3\,$\times$\,10$^{24}$\,cm$^{-2}$ and possibly much higher, with an absorption-corrected luminosity \ltwoten\,=\,(3--5)\,$\times$\,10$^{43}$\,erg\,s$^{-1}$ (Table\,2), agreeing with mid-infrared continuum and forbidden optical \oiii\ emission line indicators. The full uncertainty range on \ltwoten\ spans $\approx$\,(1--13)\,$\times$\,10$^{43}$\,erg\,s$^{-1}$ (Fig.\,\ref{fig:lumchi2}). A relatively weak neutral \fek\ emission line (EW\,$\approx$\,0.4\,keV) at 6.4\,keV is seen, together with a comparatively strong ionised Fe line consistent with 6.97\,keV emitted by \fexxvi. We explore a variety of scenarios to explain the line complex and suggest that NGC\,7674 may be similar to more powerful ULIRGs which also show similar trends of line complexity, possibly related to a high accretion rate (Section\,\ref{sec:ionised}). 

We have presented an X-ray light curve spanning 37\,years (Fig.\,\ref{fig:lc}), and find that the observed source X-ray flux has remained constant for about 20\,years, prior to which it was brighter by a factor of at least $\approx$\,10 when observed by \ginga. A past \heao\ detection was 3\,times brighter still, but background uncertainties make this measurement less reliable. A faded/switched-off AGN scenario requires a reflector size of at least 3\,pc, which is $\approx$\,30\,times larger than the dust sublimation radius of the canonical pc-scale torus in NGC\,7674, and thus is not a preferred explanation for the observed fading (Section\,\ref{sec:switchedoffscenario}). 

The alternative scenario is that a clumpy Compton-thick obscuring medium has been continuously obscuring the source for $\approx$\,21\,\p\,3\,years since the mid 1990s. Unlike known changing-look AGN, however, NGC\,7674 does not show frequent flux or spectral shape transitions, with none observed since the \sax\ observation. If a steady-state patchy obscurer does surround the nucleus, the past high state of NGC\,7674 could have represented a temporary unveiling of the nucleus (Section\,\ref{sec:obscuringscenario}). It is also noteworthy that the source has been an optical Seyfert\,2 for more than 30\,years, implying that the source has shown no evidence of being a \lq changing-look\rq\ AGN over significant periods of time in the optical. The relation between the strong nuclear outflow and the X-ray--obscuring medium, and whether the outflow is connected to the past fading, also remains to be investigated. Continued monitoring of the source will be important, as will high spatial and spectral resolution multiwavelength observations to understand these connections. 


The weakness of the neutral \fek\ emission line implies that canonical torus covering factors and Solar metallicities cannot be used in order to derive absorption correction factors. This is relevant for surveys of distant, fainter AGN where the signal to noise of the data do not allow detailed fitting of individual sources and simplifying assumptions are introduced for spectral modelling (Section\,\ref{sec:implications}). 


\section{Acknowledgements}

This research has made use of data from the \nustar\ mission, a project led by the California Institute of Technology, managed by the Jet Propulsion Laboratory, and funded by the National Aeronautics and Space Administration. We thank the \nustar\ Operations, Software, and Calibration teams for support with the execution and analysis of these observations. This research has made use of the \nustar\ Data Analysis Software (NuSTARDAS) jointly developed by the ASI Science Data Center (ASDC, Italy) and the California Institute of Technology (USA). This work made use of data supplied by the UK Swift Science Data Centre at the University of Leicester \citep{evans09_xrt}. P.G. thanks STFC for support (grant reference ST/J003697/2). 
A.C. and A.M. acknowledge support from the ASI/INAF grant I/037/12/0–011/13. A.C. acknowledges the Caltech Kingsley visitor program. We acknowledge financial support from Majlis Amanah Rakyat (MARA) Malaysia (A.A.), the Science and Technology Facilities Council (STFC) grant ST/I0015731/1 (D.M.A) and ST/K501979/1 (G.B.L). W.N.B acknowledges California Institute of Technology (Caltech) \nustar\ subcontract 44A-1092750. S.B. acknowledges financial contribution from the agreement ASI-INAF I/037/12/0. We also acknowledge NASA NuSTAR A01 Award NNX15AV27G (F.E.B.), CONICYT-Chile grants Basal-CATA PFB-06/2007 (F.E.B., C.R.), FONDECYT Regular 1141218 (F.E.B., C.R.), "EMBIGGEN" Anillo ACT1101 (F.E.B., C.R.), and the Ministry of Economy, Development, and Tourism's Millennium Science Initiative through grant IC120009, awarded to The Millennium Institute of Astrophysics, MAS (F.E.B.). We thank the referee for detailed comments which helped us to make the presentation of results clearer and to place the discussion on a more robust footing. 

\setcounter{figure}{0}
\makeatletter 
\renewcommand{\thefigure}{A\@arabic\c@figure}
\makeatother
\setcounter{section}{0}
\makeatletter 
\renewcommand{\thesection}{A\@arabic\c@section}
\makeatother
\setcounter{table}{0}
\makeatletter 
\renewcommand{\thetable}{A\@arabic\c@section}
\makeatother

\section{Appendix}

\subsection{Individual \swift\ observations}

A total of 17 \swift\ XRT observations of NGC\,7674 are available from the {\sc heasarc} archive.\footnote{{\tt http://heasarc.gsfc.nasa.gov/docs/archive.html}} Their spectra were extracted from the standard data products, and fit with simple power laws. Fixed Galactic absorption was included in these fits. In cases of insignificant detection, the gross count rate at the source position was converted to an upper limit on the observed flux assuming a power law of fixed $\Gamma$\,=\,1 because we expect the source to be obscured and display an effectively hard photon index. The resultant fits are listed in Table\,\ref{tab:xrt}. There is no evidence for significant variability amongst these measured fluxes. 

\begin{table*}
  \begin{center}
    \caption{\swift\ XRT observations of NGC\,7674.\label{tab:xrt}}
\begin{tabular}{lccccr}
\hline
Date       & ObsID       & Exposure   & Ct rate&  $\Gamma$ & Flux \\
           &             &    ks      & 10$^{-3}$\,s$^{-1}$&           & 10$^{-12}$\,erg\,s$^{-1}$\,cm$^{-2}$ \\
 (1)       &   (2)       &   (3)      &  (4)              &   (5)     &  (6) \\
\hline                                  
2011-01-28 & 00040884001 &  1492      &$<$20.0&1$^f$   & $<$\,2.15\\
2011-10-03 & 00040884002 &   845      &$<$13.8&1$^f$   & $<$\,1.49\\
2011-10-04 & 00040884003 &   845      &$<$26.7&1$^f$   & $<$\,2.87\\
2011-10-28 & 00040884004 & 4952       &5.9\,\p\,1.9&--1.52$^{+1.55}_{-u}$   & 2.13$^{+1.96}_{-1.09}$\\
2011-10-30 & 00040884005 & 3696       &7.9\,\p\,2.7&--0.23$^{+1.20}_{-1.16}$ & 1.46$^{+1.11}_{-0.66}$\\
2011-11-03 & 00040884006 & 2831       &9.2\,\p\,3.2&--0.46$^{+1.76}_{-1.68}$ & 1.63$^{+1.16}_{-0.92}$\\
2011-11-08 & 00040884007 & 2782       &4.3\,\p\,2.2&--1.11\,\p$u$ & 1.46$^{+3.01}_{-1.01}$\\
2011-11-11 & 00040884008 & 4701       &5.1\,\p\,1.9&3.15$^{+3.00}_{-2.58}$  & 0.94$^{+1.42}_{-0.36}$\\
2011-11-13 & 00040884009 & 4997       &5.8\,\p\,1.9&--0.03$^{+1.31}_{-1.32}$ & 0.98$^{+0.85}_{-0.45}$\\ 
2011-11-15 & 00040884010 & 3929       &4.4\,\p\,1.9&--1.88$^{+2.21}_{-u}$    & 2.40$^{+2.55}_{-1.60}$\\
2011-11-17 & 00040884011 &  467       &$<$25.4&1$^f$   & $<$\,2.73\\
2011-11-19 & 00040884012 & 5077       &4.0\,\p\,1.6&--0.11$^{+2.70}_{-2.67}$ & 0.99$^{+1.5}_{-0.48}$\\
2011-11-21 & 00040884013 & 4679       &3.8\,\p\,1.6&0.46$^{+1.84}_{-1.90}$ & 0.94$^{+1.15}_{-0.44}$\\ 
2013-01-29 & 00049851001 & 1454       &$<$9.5&1$^f$   & $<$\,1.02\\
2014-09-30 & 00080798001 & 1041       &$<$21.7&1$^f$   & $<$\,2.34\\
2014-10-03 & 00080798002 & 3078       &$<$9.1&1$^f$   & $<$\,0.98\\
2014-10-08 & 00080798003 & 2395       &7.4\,\p\,3.0&0.55$^{+1.79}_{-1.86}$ & 0.92$^{+1.59}_{-0.53}$\\
\hline
\end{tabular}
~\par
Ct rates (3), fitted $\Gamma$ (4) and flux (5) are for the 2--10\,keV range where the AGN is expected to dominate.\\
$^f$fixed photon indices used for flux upper limit determinations.\\
$^u$unconstrained.\\
\end{center}
\end{table*}

\subsection{Model M spectral fit}

\begin{figure*}
  \begin{center}
    \includegraphics[angle=270,width=11.5cm]{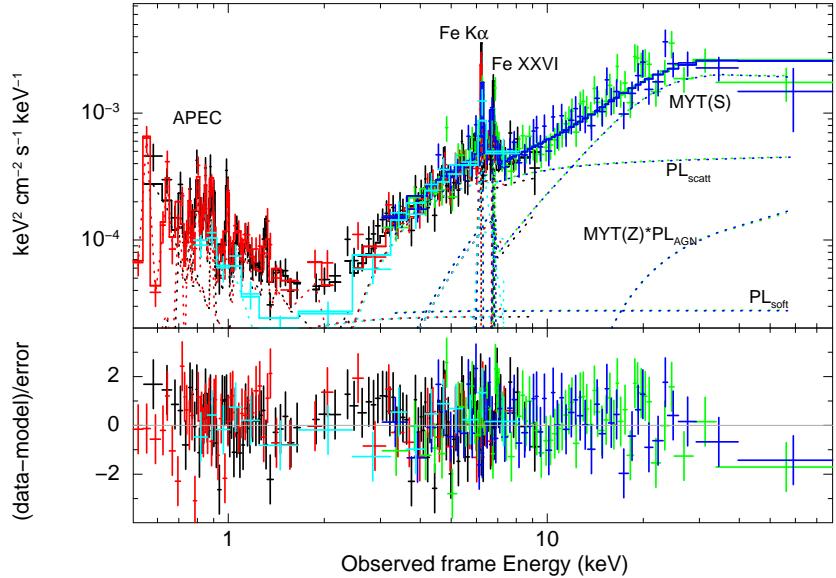}
\caption{\suzaku\ + \nustar\ data fits with the reflection model M. This is very similar to model T presented in Fig.\,\ref{fig:modelT}. The {\sc myt(z)} and {\sc myt(s)} components represent the absorption of the direct component, and the Compton-scattered component, respectively. The third \mytorus\ component is the fluorescence component producing the \fek\ and related emission features and the Compton shoulders (this would be equivalently named {\sc myt{l}}). 
 \label{fig:modelM}}
  \end{center}
\end{figure*}

Fig.\,\ref{fig:modelM} shows the fit to the default (coupled) model M incorporating \mytorus\ (\S\,\ref{sec:modelM}). The fit parameters are listed in the last data column in Table\,\ref{tab:x}. This fit is qualitatively very similar to that of model T, shown in Fig.\,\ref{fig:modelT}. 

\subsection{Black hole mass estimate from optical spectroscopy}

\begin{figure}
  \begin{center}
    \includegraphics[angle=0,width=8.5cm]{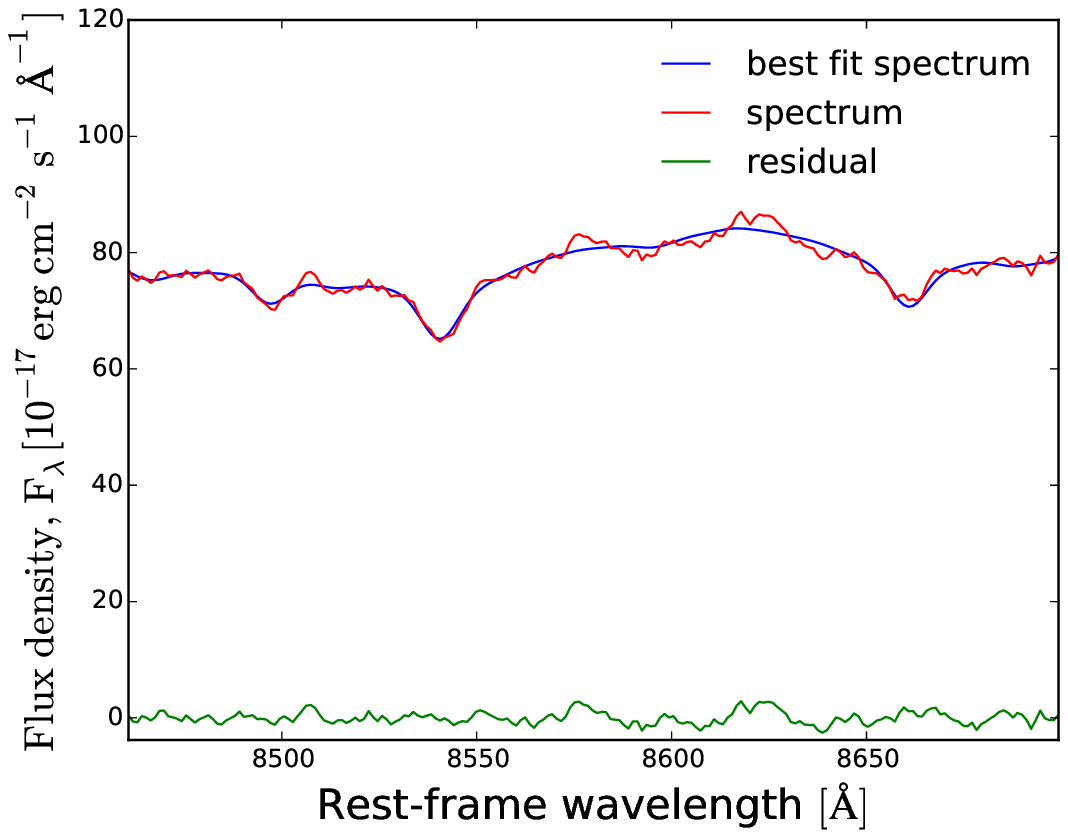}
\caption{Keck/LRIS red grism spectrum of NGC\,7674 spanning the Ca\,{\sc ii} triplet absorption feature, fitted with a pPXF algorithm. The intrinsic velocity dispersion is measured to be 91\,\p\,48\,km\,s$^{-1}$.
 \label{fig:cat}}
  \end{center}
\end{figure}

We use the penalised PiXel Fitting software \citep[{\tt pPXF},][]{ppxf} to measure the central stellar velocity dispersion ($\sigma$). For a stellar library, we used templates from the Miles Indo-U.S. Catalogue (MIUSCAT) library of stellar spectra \citep{vazdekis12} with coverage of the wavelength range of the Ca\,{\sc ii} triplet (8450--8700\,\AA).
The data and fit are presented in Fig.\,\ref{fig:cat}. The fit yielded a stellar velocity dispersion of $\sigma$\,=\,91\,\p\,48\,km\,s$^{-1}$ (1-sigma uncertainty). This value is entirely consistent with, and slightly below, the instrumental resolution $\sigma_{\rm instr.}$\,=\,107\,km\,s$^{-1}$. Measurements near the resolution limit can be unreliable, so we conservatively interpret the measurement and its uncertainty to be equivalent to an upper limit of $\sigma$\,=\,91\,+\,48\,=\,139\,km\,s$^{-1}$. Using the \Mbh--$\sigma$ relation from \citet{mcconnellma13} implies an upper limit of \Mbh\,$<$\,10$^{7.43}$\,\Msun. 

The measurement quoted by \citet{nelson95} is $\sigma$\,=\,144\,\p\,32\,km\,s$^{-1}$, which has been used to infer \Mbh\,$=$\,10$^{7.56}$\,\Msun\ in the literature \citep{bian07}. \citeauthor{nelson95} used the KPNO 2.1\,m telescope with the TI CCD, a slit width of 1\farcs 5, and 600 lines\,mm$^{-1}$ grating, observed under 1\farcs 9 seeing. The grating would have had higher spectral resolution than our LRIS observations, so it is a bit surprising that their measurement of $\sigma$ is larger than our inferred upper limit. 

However, we note that the uncertainty of 32\,km\,s$^{-1}$ suggests that the discrepancy is relatively mild. Their use of a wider slit, the bad seeing during their observation, and use of a far smaller telescope than Keck may have all resulted in lower signal-to-noise than our data. In fact, a comparison by eye of our Fig.\,\ref{fig:cat} with the spectrum presented in Fig.\,3a of \citet{nelson95} shows this to be a plausible solution to the above mismatch. A sensitive, higher resolution optical spectrum, with 1200 lines\,mm$^{-1}$, for instance, should be able to resolve this issue. 

\bibliographystyle{mnras}
\bibliography{gandhi17_rev2.bbl}

\bigskip
\noindent
{\em {\bf AUTHOR AFFILIATIONS}\newline
 $^1$Department of Physics \& Astronomy, University of Southampton, Highfield, Southampton SO17 1BJ\\
 $^2$Centre for Extragalactic Astronomy, Department of Physics, Durham University, South Road, Durham DH1 3LE\\
 $^3$Institute of Astronomy, University of Cambridge, Madingley Road, Cambridge CB3 0HA\\
 $^{4}$Jet Propulsion Laboratory, California Institute of Technology, 4800 Oak Grove Drive, Mail Stop 169-221, Pasadena, CA 91109, USA\\
  $^5$Instituto de Astrof{\'{\i}}sica and Centro de Astroingenier{\'{\i}}a, Facultad de F{\'{i}}sica, Pontificia Universidad Cat{\'{o}}lica de Chile, Casilla 306, Santiago 22, Chile\\
  $^6$Millennium Institute of Astrophysics (MAS), Nuncio Monse{\~{n}}or S{\'{o}}tero Sanz 100, Providencia, Santiago, Chile\\
  $^7$Space Science Institute, 4750 Walnut Street, Suite 205, Boulder, Colorado 80301, USA\\
  $^8$Dipartimento di Matematica e Fisica, Universit\`{a} degli Studi Roma Tre, via della Vasca Navale 84, 00146 Roma, Italy\\
  $^9$Space Sciences Laboratory, University of California, Berkeley, CA 94720, USA\\
 $^{10}$Department of Astronomy and Astrophysics, 525 Davey Lab, The Pennsylvania State University, University Park, PA 16802, USA\\
 $^{11}$Institute for Gravitation and the Cosmos, The Pennsylvania State University, University Park, PA 16802, USA\\
 $^{12}$Department of Physics, Pennsylvania State University, University Park, PA 16802, USA\\
  $^{13}$Cahill Center for Astrophysics, 1216 East California, Boulevard, California Institute of Technology, Pasadena, CA 91125, USA\\
 $^{14}$DTU Space-National Space Institute, Technical University of Denmark, Elektrovej 327, DK-2800 Lyngby, Denmark\\
  $^{15}$INAF Osservatorio Astronomico di Bologna via Ranzani 1, 40127 Bologna Italy\\
 $^{16}$Lawrence Livermore National Laboratory, Livermore, CA 94550, USA\\
 $^{17}$Max-Planck-Institut f\"{u}r Extraterrestrische Physik (MPE), Postfach 1312, D-85741 Garching, Germany\\
 $^{18}$Harvard-Smithsonian Center for Astrophysics, 60 Garden Street, Cambridge, MA 02138, USA\\
 $^{19}$Institute of Space and Astronatical Science (JAXA), 3-1-1 Yoshinodai, Sagamihara, Kanagawa, 252-5252, Japan\\
 $^{20}$European Space Astronomy Center of ESA, P.O.Box 78, Villanueva de la Ca\~{n}ada, E-28691 Madrid, Spain\\
 $^{21}$Columbia Astrophysics Laboratory, 550 W 120th Street, Columbia University, NY 10027, USA\\
 $^{22}$Institute for Astronomy, Department of Physics, ETH Zurich, Wolfgang-Pauli-Strasse 27, CH-8093 Zurich, Switzerland\\
  $^{23}$INAF Istituto di Astrofisica Spaziale e Fisica cosmica di Bologna, via Gobetti 101, I-40129, Bologna, Italy\\
  $^{24}$Dipartimento di Fisica e Astronomia (DIFA), Universit\`a di Bologna, viale Berti Pichat 6/2, 40127 Bologna, Italy\\
 $^{25}$ASDC--ASI, Via del Politecnico, 00133 Roma, Italy\\
 $^{26}$INAF--Osservatorio Astronomico di Roma, via Frascati 33, 00040 Monte Porzio Catone (RM), Italy\\
 $^{27}$X-ray Astrophysics Laboratory, NASA Goddard Space Flight Center, Greenbelt, MD 20771, USA
}

\label{lastpage}
\end{document}